\documentclass[twocolumn]{aastex62}
\usepackage[T1]{fontenc}
\usepackage{amsmath}
\usepackage{graphicx}
\usepackage{natbib}
\usepackage[scaled]{helvet}
\usepackage{epsfig}
\usepackage{url}
\bibpunct{(}{)}{;}{a}{}{,}
\usepackage{textcomp}
\usepackage{mathpazo}
\usepackage{xspace}
\usepackage{apjfonts}
\usepackage{rotating}

\usepackage{color}
\usepackage[normalem]{ulem}

\newcommand{\bjdtdb}{\ensuremath{\rm {BJD_{TDB}}}}
\newcommand{\feh}{\ensuremath{\left[{\rm Fe}/{\rm H}\right]}}

\newcommand{\teff}{\ensuremath{T_{\rm eff}}\xspace}
\newcommand{\logg}{\ensuremath{\log g}}
\newcommand\vsini{\ifmmode{v\sin{i_\star}}\else $v\sin{i_\star}$\fi}
\newcommand\sini{\ifmmode{\sin{i_\star}}\else $\sin{i_\star}$\fi}

\newcommand{\msun}{\ensuremath{\,M_\Sun}}
\newcommand{\rsun}{\ensuremath{\,R_\Sun}}
\newcommand{\lsun}{\ensuremath{\,L_\Sun}}

\newcommand{\rearth}{\ensuremath{\,R_{\rm \Earth}}\xspace}
\newcommand{\mearth}{\ensuremath{\,M_{\rm \Earth}}\xspace}
\newcommand{\re}{\ensuremath{\,R_{\rm \Earth}}\xspace}
\newcommand{\me}{\ensuremath{\,M_{\rm \Earth}}\xspace}

\newcommand{\fave}{\langle F \rangle}
\newcommand{\fluxcgs}{10$^9$ erg s$^{-1}$ cm$^{-2}$}

\newcommand{\Kepler}{{\it Kepler}}
\newcommand{\kep}{{\it Kepler}}
\newcommand{\Ktwo}{{\it K2}}
\newcommand{\tess}{{\it TESS}}
\newcommand{\tesspar}{{\it Transiting Exoplanet Survey Satellite} ({\it TESS)}}
\newcommand\mysim{\mathord{\sim}}
\newcommand{\kms}{\,km\,s$^{-1}$}
\newcommand{\ms}{\,m\,s$^{-1}$}
\newcommand{\masyr}{\ensuremath{{\rm mas\,yr}^{-1}}}

\newcommand{\thisstar}{TOI\,125\xspace}
\newcommand{\thisstarone}{TOI\,125.01\xspace}
\newcommand{\thisstartwo}{TOI\,125.02\xspace}
\newcommand{\thisstarthree}{TOI\,125.03\xspace}
\newcommand{\thisstarfour}{TOI\,125.04\xspace}
\newcommand{\thisstarfive}{TOI\,125.05\xspace}
\newcommand{\thisstarb}{TOI\,125\,b\xspace}
\newcommand{\thisstarc}{TOI\,125\,c\xspace}

\newcommand{\thistic}{TIC\,52368076\xspace}

\newcommand{\fracbrac}[2]{\left(\frac{#1}{#2}\right)}
\newcommand{\vespa}{{\texttt{vespa}}\xspace}
\newcommand{\rffig}[1]{Fig.~\ref{fig:#1}}
\newcommand{\rffigl}[1]{Figure~\ref{fig:#1}}

\newcommand{\rfsecl}[1]{\mbox{Section \ref{sec:#1}}}

\newcommand{\rftabl}[1]{Table~\ref{tab:#1}}
\newcommand{\rftabs}[2]{Tables~\ref{tab:#1} and \ref{tab:#2}}

\newcommand{\rfeql}[1]{Equation~\ref{eq:#1}}

\newcommand{\cfa}{Center for Astrophysics | Harvard \& Smithsonian, 60 Garden St, Cambridge, MA 02138, USA}
\newcommand{\umich}{Astronomy Department, University of Michigan, 1085 S University Avenue, Ann Arbor, MI 48109, USA}
\newcommand{\utaustin}{Department of Astronomy, The University of Texas at Austin, Austin, TX 78712, USA}
\newcommand{\MIT}{Department of Physics and Kavli Institute for Astrophysics and Space Research, Massachusetts Institute of Technology, Cambridge, MA 02139, USA}
\newcommand{\eaps}{Department of Earth, Atmospheric, and Planetary Sciences, Massachusetts Institute of Technology, 77 Massachusetts Avenue, Cambridge, MA 02139, USA}
\newcommand{\mitaero}{Department of Aeronautics and Astronautics, Massachusetts Institute of Technology, 125 Massachusetts Ave, Cambridge, MA 02139, USA}
\newcommand{\uflorida}{Department of Astronomy, University of Florida, 211 Bryant Space Science Center, Gainesville, FL, 32611, USA}
\newcommand{\riverside}{Department of Earth Sciences, University of California,
Riverside, CA 92521, USA}
\newcommand{\usq}{Centre for Astrophysics, University of Southern Queensland, West Street, Toowoomba, QLD 4350, Australia}
\newcommand{\ames}{NASA Ames Research Center, Moffett Field, CA, 94035, USA}
\newcommand{\geneva}{Observatoire de l'Universit\'e de Gen\`eve, 51 chemin des Maillettes,
1290 Versoix, Switzerland}
\newcommand{\uw}{Astronomy Department, University of Washington, Seattle, WA 98195 USA}
\newcommand{\warwick}{Deptartment of Physics, University of Warwick, Gibbet Hill Road, Coventry CV4 7AL, UK}
\newcommand{\warwickceh}{Centre for Exoplanets and Habitability, University of Warwick, Gibbet Hill Road, Coventry CV4 7AL, UK}
\newcommand{\princeton}{Department of Astrophysical Sciences, Princeton University, 4 Ivy Lane, Princeton, NJ, 08544, USA}
\newcommand{\liegestar}{STAR Institute, Universit\'e de Li\`ege, All\'ee du 6 Ao\^ut 17, 4000 Li\`ege, Belgium}
\newcommand{\liegeur}{UR Astrobiology, Universit\'e de Li\`ege, 19C All\'ee du 6 Ao\^ut, 4000 Li\`ege, Belgium}
\newcommand{\vanderbilt}{Department of Physics and Astronomy, Vanderbilt University, Nashville, TN 37235, USA}
\newcommand{\fisk}{Department of Physics, Fisk University, 1000 17th Avenue North, Nashville, TN 37208, USA}

\newcommand{\toronto}{Dunlap Institute for Astronomy and Astrophysics, University of Toronto, Ontario M5S 3H4, Canada}
\newcommand{\unc}{Department of Physics and Astronomy, University of North Carolina at Chapel Hill, Chapel Hill, NC 27599, USA}
\newcommand{\iac}{Instituto de Astrof\'isica de Canarias (IAC), E-38205 La Laguna, Tenerife, Spain}
\newcommand{\lalaguna}{Departamento de Astrof\'isica, Universidad de La Laguna (ULL), E-38206 La Laguna, Tenerife, Spain}
\newcommand{\louisville}{Department of Physics and Astronomy, University of Louisville, Louisville, KY 40292, USA}
\newcommand{\aavso}{American Association of Variable Star Observers, 49 Bay State Road, Cambridge, MA 02138, USA}
\newcommand{\utokyo}{Department of Astronomy, The University of Tokyo, 7-3-1 Hongo, Bunky\={o}, Tokyo 113-8654, Japan}
\newcommand{\naoj}{National Astronomical Observatory of Japan, 2-21-1 Osawa, Mitaka, Tokyo 181-8588, Japan}
\newcommand{\jstpresto}{JST, PRESTO, 7-3-1 Hongo, Bunkyo-ku, Tokyo 113-0033, Japan}
\newcommand{\astrobiojapan}{Astrobiology Center, 2-21-1 Osawa, Mitaka, Tokyo 181-8588, Japan}

\newcommand{\nexsci}{Caltech/IPAC -- NASA Exoplanet Science Institute 1200 E. California Ave, Pasadena, CA 91125, USA}
\newcommand{\ucsc}{Department of Astronomy and Astrophysics, University of
California, Santa Cruz, CA 95064, USA}
\newcommand{\gsfc}{Exoplanets and Stellar Astrophysics Laboratory, Code 667, NASA Goddard Space Flight Center, Greenbelt, MD 20771, USA}
\newcommand{\sgtinc}{SGT, Inc./NASA AMES Research Center, Mailstop 269-3, Bldg T35C, P.O. Box 1, Moffett Field, CA 94035, USA}

\newcommand{\brorfelde}{Brorfelde Observatory, Observator Gyldenkernes Vej 7, DK-4340 T\o{}ll\o{}se, Denmark}
\newcommand{\noqsi}{Noqsi Aerospace, Ltd., 15 Blanchard Avenue, Billerica, Massachusetts 01821, USA}
\newcommand{\protologic}{Proto-Logic Consulting LLC, Washington, DC 20009, USA}
\newcommand{\scsu}{Department of Physics, Southern Connecticut State University, 501 Crescent St., New Haven, CT 06515, USA}
\newcommand{\telaviv}{School of Physics and Astronomy, Raymond and Beverly Sackler Faculty of Exact Sciences, Tel Aviv University, Tel Aviv 6997801, Israel}
\newcommand{\dtuspace}{DTU Space, National Space Institute, Technical University of Denmark, Elektrovej 327, DK-2800 Lyngby, Denmark}
\newcommand{\seti}{SETI Institute, 189 Bernardo Ave., Suite 200, Mountain View, CA 94043, USA}
\newcommand{\swarthmore}{Deptartment of Physics \& Astronomy, Swarthmore College, Swarthmore PA 19081, USA}
\newcommand{\cadiayyad}{Oukaimeden Observatory, High Energy Physics and Astrophysics Laboratory, Cadi Ayyad University, Marrakech, Morocco}
\newcommand{\cavendish}{Cavendish Laboratory, University of Cambridge, JJ Thomson Avenue, Cambridge, CB3 0H3, UK}

\newcommand{\torres}{\altaffiliation{Juan Carlos Torres Fellow}}
\newcommand{\sagan}{\altaffiliation{NASA Sagan Fellow}}
\newcommand{\hubble}{\altaffiliation{NASA Hubble Fellow}}
\newcommand{\nsfgrfp}{\altaffiliation{NSF Graduate Research Fellow}}
\newcommand{\kavli}{\altaffiliation{Kavli Fellow}}

\begin{document}

\title{Near-resonance in a system of sub-Neptunes from {\it TESS}}
\AuthorCallLimit=15
\correspondingauthor{Samuel N. Quinn}
\email{squinn@cfa.harvard.edu}

\author[0000-0002-8964-8377]{Samuel N. Quinn}
\affiliation{\cfa}

\author[0000-0002-7733-4522]{Juliette C. Becker}
\affiliation{\umich}

\author[0000-0001-8812-0565]{Joseph E. Rodriguez}
\affiliation{\cfa}

\author[0000-0002-1032-0783]{Sam Hadden}
\affiliation{\cfa}

\author[0000-0003-0918-7484]{Chelsea X. Huang}
\torres
\affiliation{\MIT}

\author[0000-0002-8537-5711]{Timothy D. Morton}
\affiliation{\uflorida}


\author[0000-0002-8167-1767]{Fred Adams} 
\affiliation{\umich}

\author[0000-0002-5080-4117]{David Armstrong} 
\affiliation{\warwick}
\affiliation{\warwickceh}

\author[0000-0003-3773-5142]{Jason D. Eastman} 
\affiliation{\cfa}

\author[0000-0002-1160-7970]{Jonathan Horner} 
\affiliation{\usq}

\author[0000-0002-7084-0529]{Stephen R. Kane} 
\affiliation{\riverside}

\author{Jack J.\ Lissauer} 
\affiliation{\ames}

\author[0000-0002-6778-7552]{Joseph D. Twicken} 
\affiliation{\seti}

\author[0000-0001-7246-5438]{Andrew Vanderburg}
\sagan
\affiliation{\utaustin}

\author[0000-0001-9957-9304]{Rob Wittenmyer} 
\affiliation{\usq}


\author{George R. Ricker}
\affiliation{\MIT}

\author[0000-0001-6763-6562]{Roland K. Vanderspek}
\affiliation{\MIT}

\author[0000-0001-9911-7388]{David W. Latham}
\affiliation{\cfa}

\author{Sara Seager}
\affiliation{\MIT}
\affiliation{\eaps}
\affiliation{\mitaero}

\author[0000-0002-4265-047X]{Joshua N.~Winn}
\affiliation{\princeton}

\author[0000-0002-4715-9460]{Jon M.~Jenkins}
\affiliation{\ames}


\author[0000-0002-0802-9145]{Eric Agol} 
\affiliation{\uw}

\author{Khalid Barkaoui}
\affiliation{\cadiayyad}
\affiliation{\liegeur}

\author[0000-0002-5627-5471]{Charles A.\ Beichman} 
\affiliation{\nexsci}

\author{Fran\c{c}ois Bouchy} 
\affiliation{\geneva}

\author[0000-0002-0514-5538]{L. G. Bouma} 
\affiliation{\princeton}

\author[0000-0001-9892-2406]{Artem Burdanov}
\affiliation{\liegestar}

\author{Jennifer Campbell} 
\altaffiliation{formerly Wyle Labs/NASA Ames Research Center}
\noaffiliation{}

\author{Roberto Carlino} 
\affiliation{\sgtinc}

\author{Scott M. Cartwright} 
\affiliation{\protologic}

\author[0000-0002-9003-484X]{David Charbonneau} 
\affiliation{\cfa}

\author[0000-0002-8035-4778]{Jessie L. Christiansen} 
\affiliation{\nexsci}

\author[0000-0002-5741-3047]{David Ciardi} 
\affiliation{\nexsci}

\author[0000-0001-6588-9574]{Karen A.\ Collins} 
\affiliation{\cfa}

\author[0000-0003-2781-3207]{Kevin I.\ Collins} 
\affiliation{\vanderbilt}

\author[0000-0003-2239-0567]{Dennis M.\ Conti} 
\affiliation{\aavso}

\author{Ian J.\ M.\ Crossfield} 
\affiliation{\MIT}

\author[0000-0002-6939-9211]{Tansu Daylan} 
\kavli
\affiliation{\MIT}

\author[0000-0001-7730-2240]{Jason Dittmann} 
\affiliation{\MIT}

\author[0000-0003-2996-8421]{John Doty} 
\affiliation{\noqsi}

\author[0000-0003-2313-467X]{Diana Dragomir} 
\hubble
\affiliation{\MIT}

\author{Elsa Ducrot}
\affiliation{\liegeur}

\author{Michael Gillon}
\affiliation{\liegeur}

\author[0000-0002-5322-2315]{Ana Glidden} 
\affiliation{\MIT}
\affiliation{\eaps}

\author[0000-0003-1748-5975]{Robert F. Goeke} 
\affiliation{\MIT}

\author{Erica J.\ Gonzales} 
\nsfgrfp
\affiliation{\ucsc}

\author[0000-0002-7650-3603]{Krzysztof G.\ He{\l}miniak} 
\affiliation{Nicolaus Copernicus Astronomical Center, Polish Academy of Sciences, ul. Rabia\'{n}ska 8, 87-100 Toru\'{n}, Poland}

\author{Elliott P. Horch}
\affiliation{\scsu}

\author[0000-0002-2532-2853]{Steve B.\ Howell} 
\affiliation{\ames}

\author[0000-0001-8923-488X]{Emmanuel Jehin}
\affiliation{\liegestar}

\author[0000-0002-4625-7333]{Eric L. N. Jensen}
\affiliation{\swarthmore}

\author[0000-0003-0497-2651]{John F.\ Kielkopf} 
\affiliation{\louisville}

\author[0000-0002-2607-138X]{Martti H. Kristiansen}
\affiliation{\dtuspace}
\affiliation{\brorfelde}

\author[0000-0001-9380-6457]{Nicholas Law} 
\affiliation{\unc}

\author[0000-0003-3654-1602]{Andrew W. Mann} 
\affiliation{\unc}

\author{Maxime Marmier} 
\affiliation{\geneva}

\author[0000-0001-7233-7508]{Rachel A.\ Matson} 
\affiliation{\ames}

\author{Elisabeth Matthews} 
\affiliation{\MIT}

\author{Tsevi Mazeh} 
\affiliation{\telaviv}

\author{Mayuko Mori} 
\affiliation{\utokyo}

\author[0000-0001-9087-1245]{Felipe Murgas} 
\affiliation{\iac}
\affiliation{\lalaguna}

\author{Catriona Murray}
\affiliation{\cavendish}

\author[0000-0001-8511-2981]{Norio Narita} 
\affiliation{\utokyo}
\affiliation{\jstpresto}
\affiliation{\astrobiojapan}
\affiliation{\naoj}
\affiliation{\iac}

\author[0000-0002-5254-2499]{Louise D. Nielsen} 
\affiliation{\geneva}

\author[0000-0001-9305-9631]{Ga\"el Ottoni} 
\affiliation{\geneva}

\author{Enric Palle} 
\affiliation{\iac}
\affiliation{\lalaguna}

\author{Rafa{\l} Paw{\l}aszek} 
\affiliation{Nicolaus Copernicus Astronomical Center, Polish Academy of Sciences, ul. Rabia\'{n}ska 8, 87-100 Toru\'{n}, Poland}

\author{Francesco Pepe} 
\affiliation{\geneva}

\author{Jerome Pitogo de Leon} 
\affiliation{\utokyo}

\author[0000-0003-1572-7707]{Francisco J. Pozuelos} 
\affiliation{\liegestar}
\affiliation{\liegeur}

\author{Howard M. Relles} 
\affiliation{\cfa}

\author{Joshua E.\ Schlieder} 
\affiliation{\gsfc}

\author{Daniel Sebastian}
\affiliation{\liegeur}

\author{Damien S\'egransan} 
\affiliation{\geneva}

\author[0000-0002-1836-3120]{Avi Shporer} 
\affiliation{\MIT}

\author[0000-0002-3481-9052]{Keivan G.\ Stassun} 
\affiliation{\vanderbilt}
\affiliation{\fisk}

\author[0000-0002-6510-0681]{Motohide Tamura} 
\affiliation{\utokyo}
\affiliation{\astrobiojapan}
\affiliation{\naoj}


\author{St\'ephane Udry} 
\affiliation{\geneva}

\author[0000-0002-3249-3538]{Ian Waite} 
\affiliation{\usq}

\author[0000-0002-0619-7639]{Carl Ziegler} 
\affiliation{\toronto}

\shorttitle{Sub-Neptunes orbiting \thisstar}
\shortauthors{Quinn et al.}

\begin{abstract}
We report the \tesspar\ detection of a multi-planet system orbiting the $V=10.9$ K0 dwarf \thisstar. We find evidence for up to five planets, with varying confidence. Three high signal-to-noise transit signals correspond to sub-Neptune-sized planets ($2.76$, $2.79$, and $2.94$\ \rearth), and we statistically validate the planetary nature of the two inner planets ($P_b = 4.65$\ days, $P_c = 9.15$ days). With only two transits observed, we report the outer object ($P_{.03} = 19.98$\ days) as a high signal-to-noise ratio planet candidate. We also detect a candidate transiting super-Earth ($1.4$\,\rearth) with an orbital period of only $12.7$\ hours and a candidate Neptune-sized planet ($4.2$\,\rearth) with a period of $13.28$\ days, both at low signal-to-noise. This system is amenable to mass determination via radial velocities and transit timing variations, and provides an opportunity to study planets of similar size while controlling for age and environment. The ratio of orbital periods between \thisstarb and c ($P_c/P_b = 1.97$) is slightly smaller than an exact 2:1 commensurability and is atypical of multiple planet systems from \Kepler, which show a preference for period ratios just {\it wide} of first-order period ratios. A dynamical analysis refines the allowed parameter space through stability arguments and suggests that, despite the nearly commensurate periods, the system is unlikely to be in resonance. 
\end{abstract}


\section{Introduction} 

NASA's \tesspar\ \citep{ricker:2015} is a (nearly) all-sky survey, the primary objective of which is to discover and characterize transiting planets smaller than Neptune orbiting the nearest and brightest stars in the sky. While the space-based transit survey carried out by \Kepler\ \citep{borucki:2010} led to breakthroughs in our understanding of the occurrence rates of planetary systems \citep[e.g.,][]{Fressin:2013} and the various dynamical configurations of multi-planet systems \citep[e.g.,][]{Lissauer:2011b,Fabrycky:2014}, \tess\ is designed to discover the planetary systems most amenable to detailed characterization through follow-up observations. Typical \tess\ planet hosts will be several magnitudes brighter than those from \Kepler\ \citep{Sullivan:2015,barclay:2018,huang:2018a}, and these statistically rare systems are amenable to the most precise radial-velocity (RV) mass measurements and more efficient atmospheric characterization. This expectaction is borne out by the experience of the \Ktwo\ mission \citep{Howell:2014}, which surveyed a larger area of sky than \Kepler\ and discovered a number of bright planetary systems like the ones expected from \tess\ \citep[see, e.g.,][ and the overview presented therein]{Rodriguez:2018a}. \tess\ should also detect objects that are {\it intrinsically} rare, such as events occurring on astronomically short timescales, or the unlikely outcomes of dynamical interactions. Further study of these benchmark objects may lead to breakthroughs in our understanding of the fundamental processes that govern the formation and evolution of planetary systems.

Data from just the first \tess\ observing sector ($27.4$\ days, or two spacecraft orbits) have already begun to fulfill this promise. A $2.1$-$\rearth$\ planet transiting the 5th magnitude star $\pi$\ Mensae already has a mass measurement ($4.8\ \mearth$) because the star was previously known to host a long-period giant planet and there exist extensive archival RV measurements \citep{huang:2018b,gandolfi:2018}. The star is one of the very brightest known to host a transiting planet, which will enable further detailed characterization. A second system \citep[LHS 3844;][]{vanderspek:2018} is an Ultra-Short Period (USP) Earth-sized planet ($R_p = 1.32\ \rearth$) in an $11$-hour orbit around a late M dwarf $15$\ pc away. It is one of the nearest planetary systems, and in many respects the USP planet that is most amenable to follow-up studies. The remaining $25$\ observing sectors in the two-year prime \tess\ mission will survey additional bright stars, some for longer periods of time, and will lead to the discovery of many more benchmark planetary systems.

Among the myriad discoveries from the \Kepler\ mission, the detection of systems of multiple transiting planets and the subsequent study of their ensemble properties remain among the most impactful results \citep[e.g.,][]{steffen:2010,lissauer:2011a,latham:2011}. Multi-planet transiting systems allow investigations of formation and evolution processes through measurements of mutual inclinations, adjacent planet sizes, planet spacings, stellar obliquities, mass measurements via transit timing variations, and more. Given the prevalence of multi-transiting systems, \tess\ will build upon the \Kepler\ legacy by discovering the nearest and brightest such systems, as well as the rare examples.

\begin{figure}[!t]
\centering\includegraphics[width=\linewidth]{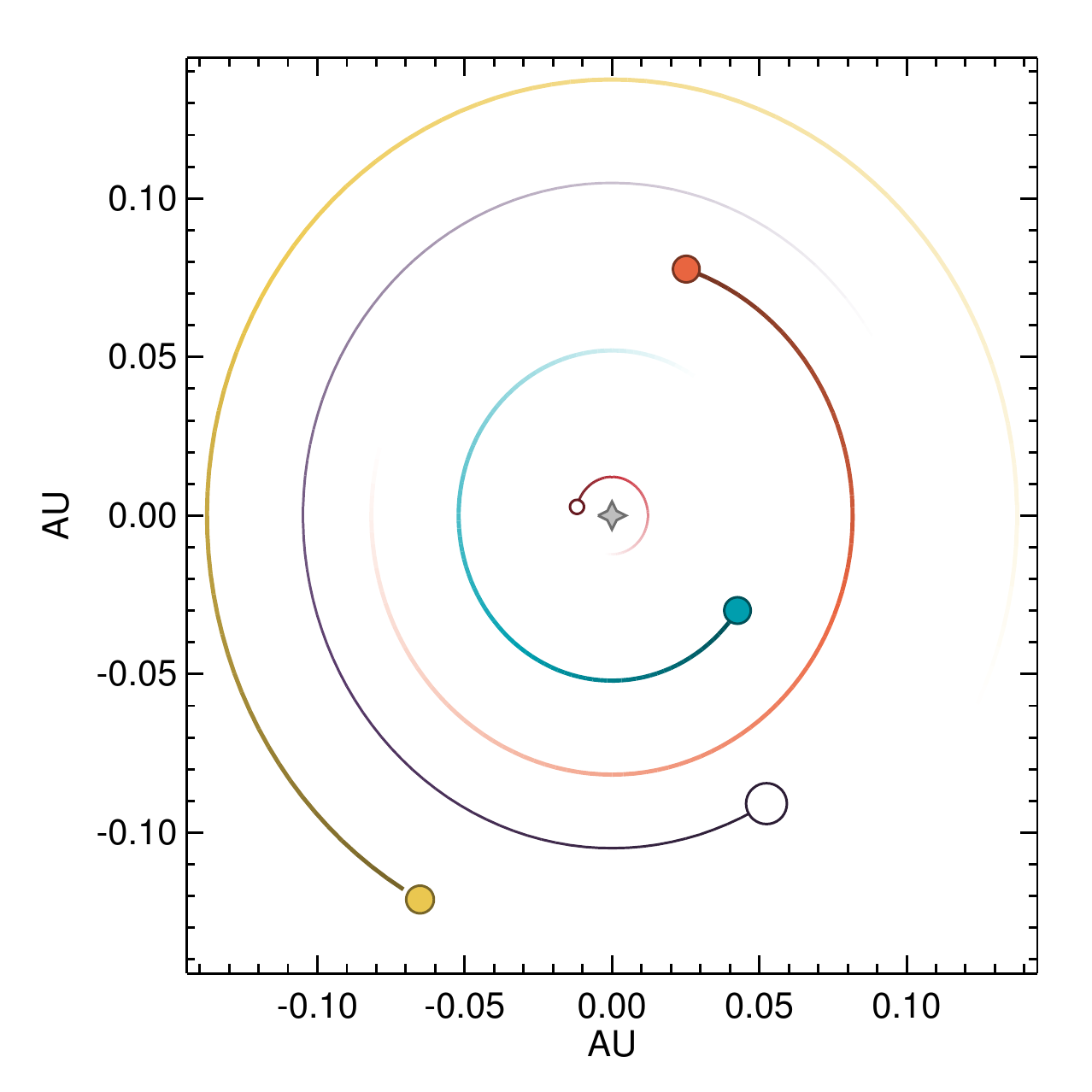}
\caption{A top-down view of the \thisstar\ planetary system. The planet sizes are drawn to scale relative to each other. The low-SNR candidates (\thisstarfour\ and .05) are shown as open circles, while the high-SNR candidate and validated planets are filled circles. We note that the derived size of .05 is very uncertain because its transit is grazing. Moreover, planets like \thisstarfive\ are a priori likely to be small; if real, its true size is probably smaller than shown.}
\label{fig:arch}
\end{figure}

In this paper, we present the discovery and validation of a system of multiple transiting planets orbiting the star \thistic, which has been assigned \tess\ Object of Interest (TOI) number \thisstar. The proposed architecture of the system is illustrated in \rffigl{arch}. We identify three candidates with high signal-to-noise (SNR) transits (filled circles), as well as two low-SNR candidates (open circles), one of which is an ultra-short-period terrestrial candidate.

We describe the analysis of data from \tess, ground-based follow-up, and archival imaging in \rfsecl{data}, fit a global model to all available data in \rfsecl{exofast}, present a statistical validation of the planets in \rfsecl{vespa}, investigate the dynamics in \rfsecl{dynamics}, and discuss the properties of the system and prospects for future characterization in \rfsecl{discussion}.

\section{Data \& Analysis}
\label{sec:data}

\subsection{TESS Photometry}
\label{sec:tess}

\begin{figure*}[!t]
\centering\includegraphics[width=\linewidth]{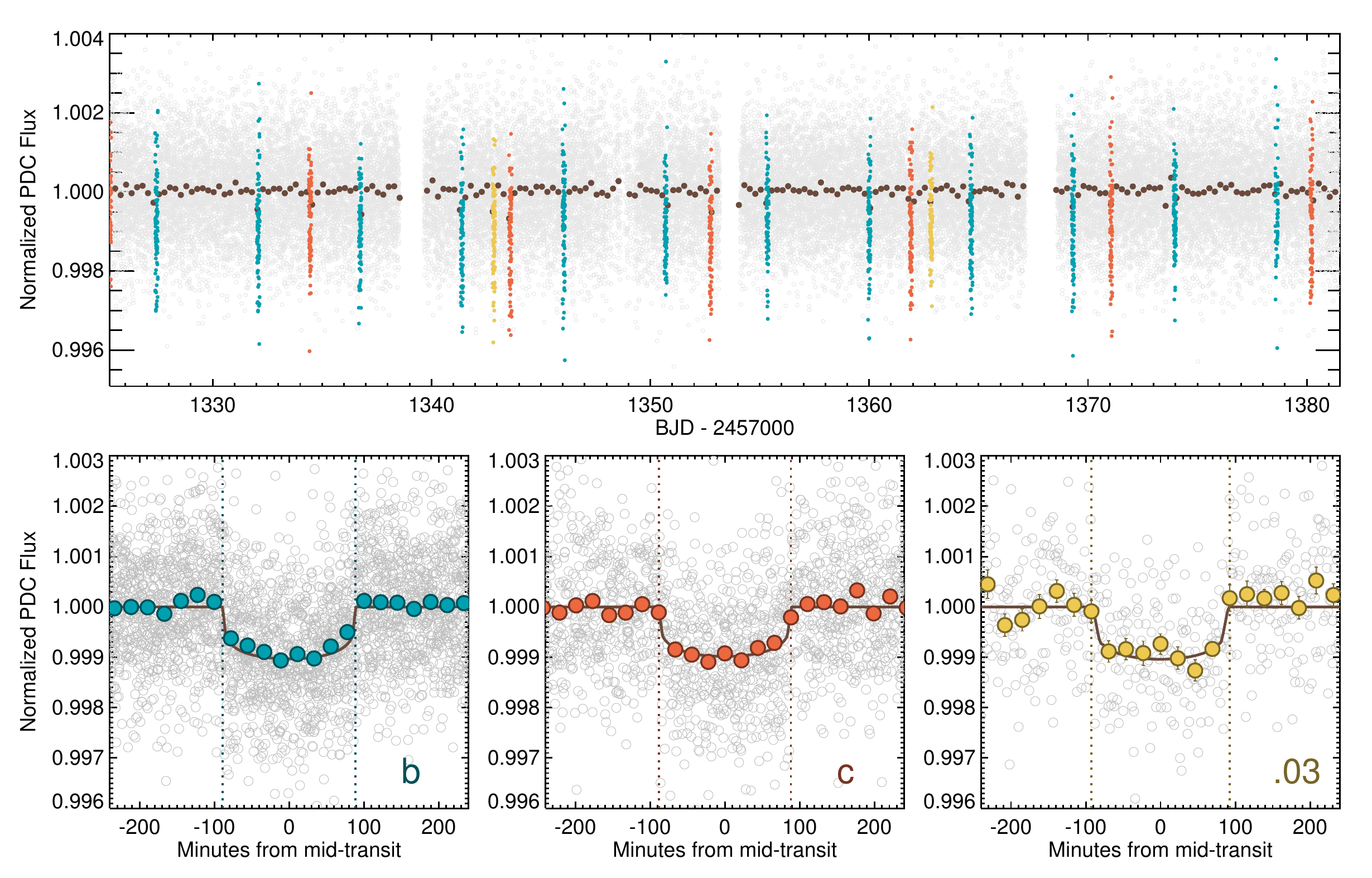}
\caption{(Top) The full \tess\ light curve of \thisstar from Sectors 1 and 2. The lightcurve has been flattened using the technique from \citet{Vanderburg:2016b}. We show the individual two-minute cadence measurements (open gray circles) and the same data in six-hour bins (brown circles). In-transit cadences corresponding to the inner, middle, and outer planets are plotted with blue, orange, and yellow colored circles, respectively. (Bottom) The phase-folded transits of each planet, with individual observations (open gray circles) and binned data (filled colored circles, chosen to have eight bins per transit duration). The best fit EXOFASTv2 models are plotted in brown. Vertical dotted lines indicate the full transit durations.}
\label{fig:lc}
\end{figure*}

\thistic\ was observed by \tess\ in the first two sectors of the prime mission (2018 Jul 25 through 2018 Sep 21), on CCD 1 of Camera 3 in Sector 1 and CCD 2 of Camera 3 in Sector 2. The CCDs obtain images at a two-second cadence, which are summed on board the spacecraft to produce images with the appropriate effective exposure time. All stars within the \tess\ field of view are observed with an effective exposure time of $30$\ minutes, but a subset of stars (including \thistic) were pre-selected, primarily on the basis of planet detectability \citep{stassun:2018b}, for data to also be returned to Earth at a two-minute cadence. 

The two-minute data were reduced with the Science Processing Operations Center (SPOC) pipeline \citep{Jenkins:2015,Jenkins:2016}, adapted from the pipeline for the {\it Kepler} mission at the NASA
Ames Research Center \citep{jenkins:2010}. Two transit signals were strongly detected with periods of $4.65$\ and $9.15$\ days and signal-to-noise ratios (SNR) of $20.1$\ and $16.4$\, respectively. These candidates were assigned identifiers  \thisstarone and \thisstartwo\ by the \tess\ team. 
An additional signal, \thisstarthree, was detected with only two transit-like events at a period of $19.98$\ days and SNR of $9.8$. In the analyses that follow, we use the Pre-search Data Conditioning (PDC) light curve from SPOC \citep[see][]{stumpe:2012}. \rffigl{lc} shows the PDC light curve after flattening (we note that the raw PDC light curve looks nearly identical, as the star is photometrically very quiet). Interruptions in data acquisition occur at the perigee of each \tess\ orbit (once every 13.7 days) and last approximately 1 day, during which time the spacecraft reorients to downlink data. The second orbit of Sector 1 included a two-day period during which the data were of lesser quality due to a one-time occurrence of abnormally unstable spacecraft pointing. The worst of these data were flagged by SPOC and removed, which can be seen as an under-density of points in \rffigl{lc} just before BTJD 1350. Fortuitously, none of the transits of these three candidates occurred at this time. During Sectors 1 and 2, the spacecraft thrusters were fired periodically (approximately every 2.5 days) to reduce the speed of the reaction wheels, allowing them to operate at frequencies that introduced less pointing jitter. In $10$- to $15$-minute intervals around these "momentum dumps", we removed data from our analysis.

\subsection{Additional Planet Candidates}
\label{moreplanets}

\begin{figure}[!t]
\centering\includegraphics[width=\linewidth]{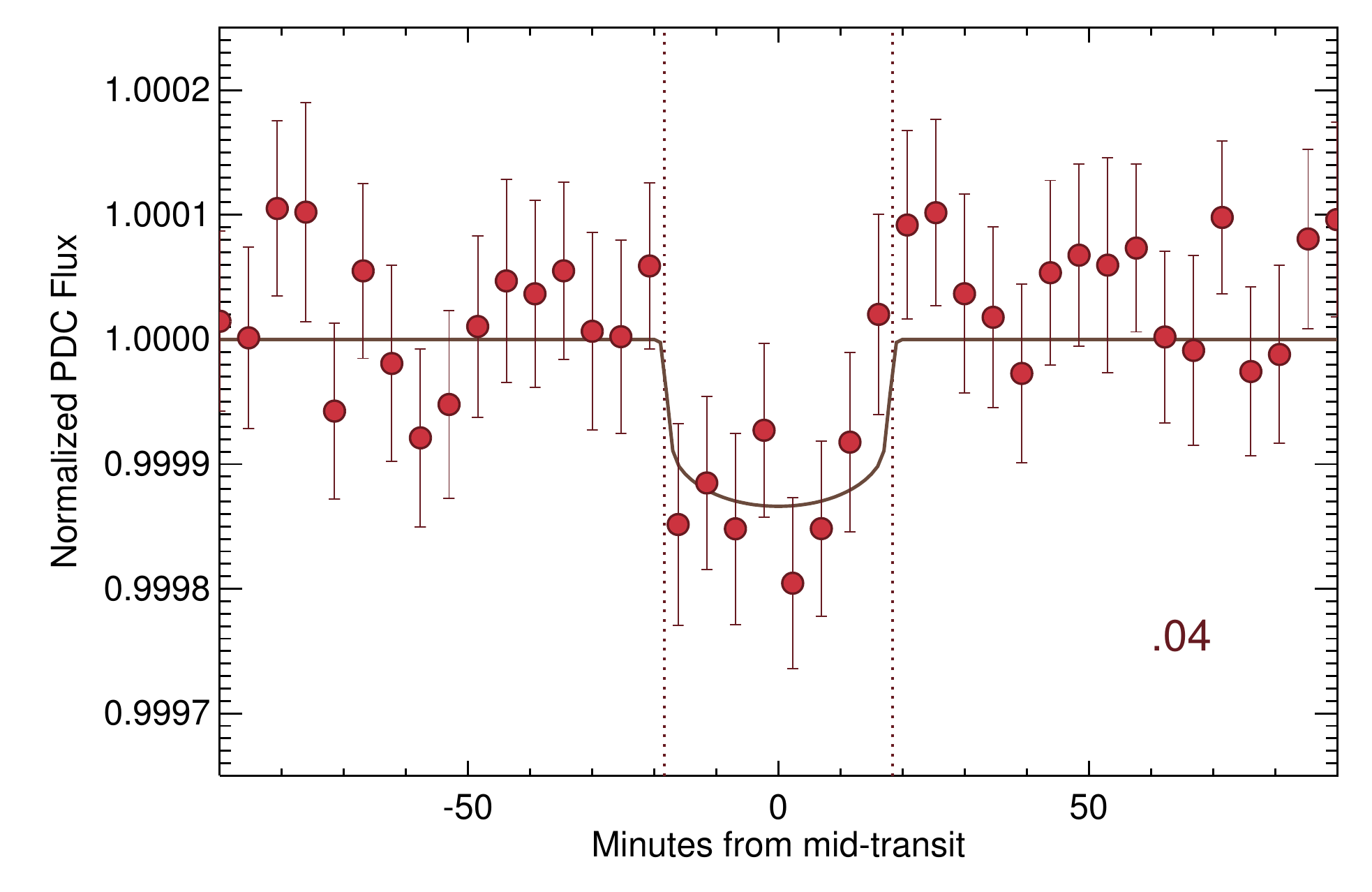}
\caption{The phase-folded transit of the candidate ultra short period super Earth, \thisstarfour. We plot the binned photometry (filled red circles), as the individual two-minute data extend far beyond the y-axis range. The best fit EXOFASTv2 model is plotted in brown. Vertical dotted lines indicate the full transit duration.}
\label{fig:usp}
\end{figure}

\begin{figure}[!t]
\centering\includegraphics[width=\linewidth]{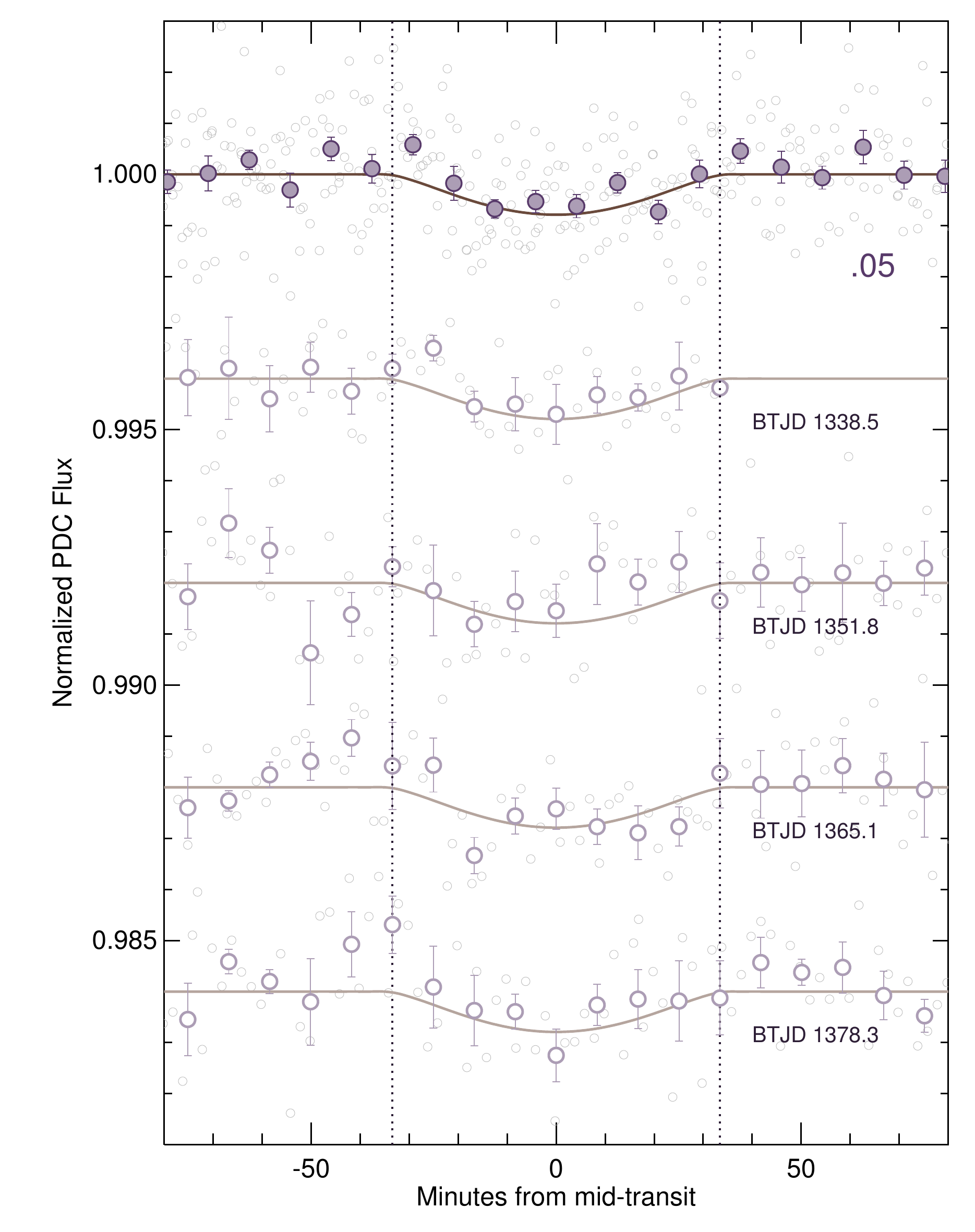}\\[-3ex]
\centering\includegraphics[width=\linewidth]{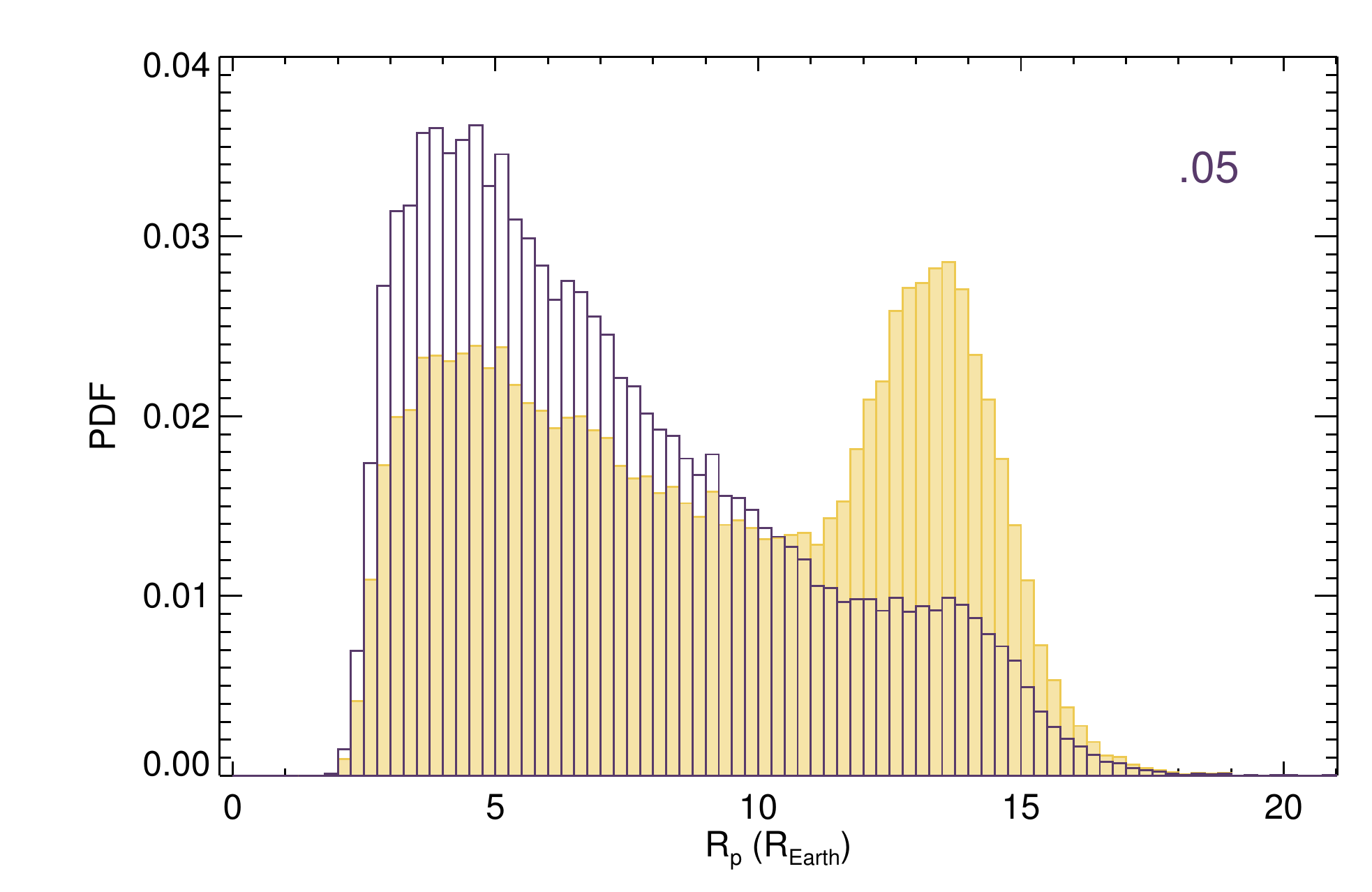}
\caption{{\it Top}: The phase-folded and individual transits of the low-SNR candidate \thisstarfive. {\it Bottom}: The EXOFASTv2 marginalized posterior distribution for $R_{\rm .05}$ (filled yellow bars), and the revised distribution when constrained by the CORALIE RVs and a mass-radius relationship (open purple bars).}
\label{fig:cand}
\end{figure}

Following the convention from \kep, we adopt a formal significance threshold of $7.1 \sigma$, and the three candidate planets described above are the only formally significant periodic signals in the data. However, we do detect lower SNR transit-like signals at two other periods. The first, \thisstarfour, has a period of $0.52854$\ days and a depth of $180$\ ppm ($5.2$-$\sigma$), which corresponds to a planet radius of $1.36$\,\rearth (see \rffigl{usp}). The second, \thisstarfive, was detected at a period of $13.2780$\,days and a depth of $675$\ ppm ($5.1$-$\sigma$). If this were a central transit, it would correspond to a planet radius of $\mysim2.2$\,\rearth, but it is best modeled as a grazing transit (see \rffigl{cand}), so when unconstrained by other data even giant planets are allowed. However, our CORALIE spectroscopy (\rfsecl{recon}) rules out the largest companions, and we ultimately derive a radius of $\mysim4.2$\,\rearth, still with large error bars. Given the low SNR of these candidates and the non-Gaussian noise of the first \tess\ Sectors, we do not consider the signals strong enough to be validated as planets, particularly \thisstarfive, which only shows four transits of varying quality. Nevertheless, the presence of three other strong planet candidates makes these signals more intriguing, and we note them here so that they can be taken into account during follow-up observations and subsequent analysis of the first three candidates.

\subsection{Ground-based Photometry}
\label{sec:sg1}

\begin{figure*}[!t]
\centering\includegraphics[width=\linewidth]{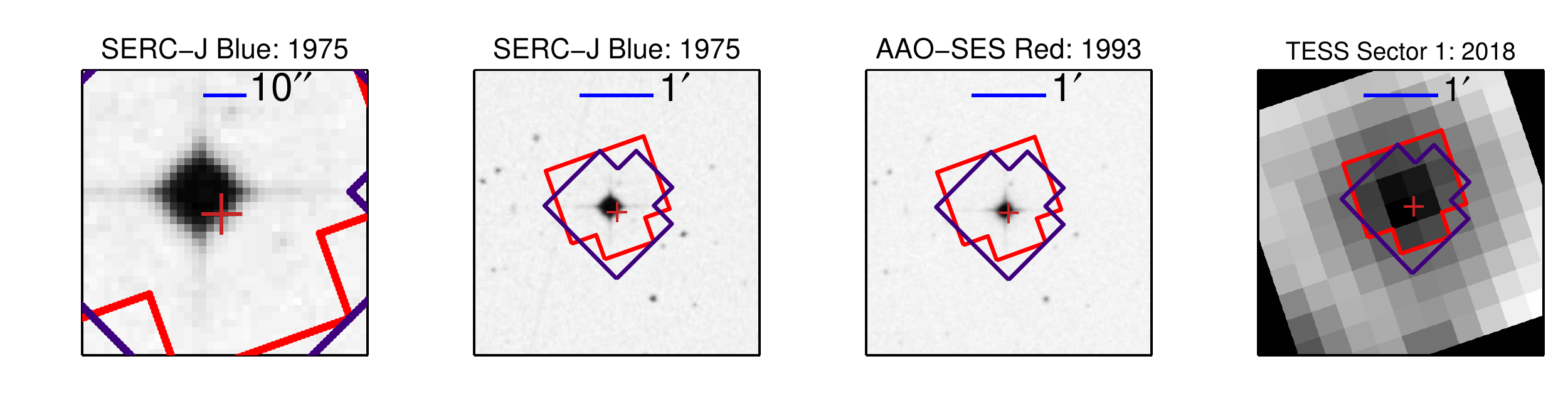}
\caption{Archival images of \thisstar\ (left 3 panels) and the \tess\ image from Sector 1 (right). Photometric apertures used in Sectors 1 (red outline) and 2 (blue outline) are also shown. The proper motion of the star has led to motion of $\mysim7\farcs$4\ in 43 years. Its current location (red cross) is marked in all images, and we detect no background sources at this location in previous epochs.}
\label{fig:patient}
\end{figure*}

Given the $21\arcsec$\ \tess\ pixels and a PSF a few pixels wide, the light from an individual star on the detector can extend well beyond $1\arcmin$. In order to capture most of the light from the target star, the \tess\ photometric apertures must also be large. Therefore, even apparently isolated stars may be contaminated by relatively distant neighbors, with the exact contamination fraction depending upon the aperture choice and the magnitudes of the stars (see, for example, the size of the PSF in the \tess\ image of \thisstar\ shown in \rffigl{patient}). If a neighboring star is an eclipsing binary (EB), deep eclipses can be diluted to resemble shallow planetary transits. While previous experience with \kep\ candidate multi-planet systems shows the vast majority to be real planets \citep[e.g.,][]{lissauer:2012}, the larger \tess\ pixels and aperture create more opportunity for unassociated EBs to contaminate real planetary systems (producing candidate multi-planet systems containing both real planets and false positives). Centroid analysis of the \tess\ difference images (comparing the in-transit to out-of-transit flux) is often effective at identifying nearby EBs, but transit signals with a small number of events or contaminants within about a pixel might not be robustly detected. We therefore observed \thisstar\ with ground-based facilities at predicted times of transit to search for deep eclipses in nearby stars. We enumerate these observations in  \rftabl{sg1_table}. In order to produce the $1$\ mmag events of \thisstarb, c, or .03 with even a $50$\% eclipse, a nearby star must be no more than $6.9$\ magnitudes fainter (and fully blended in the TESS aperture). Among {\it Gaia} DR2 sources, only one nearby star is bright enough, but at a distance of $75\arcsec$\ (to the SSW), it is not fully blended within the TESS aperture (see \rffigl{patient}). Nonetheless, we search this and other nearby stars for evidence of deep eclipses and we find no indication of contamination from nearby EBs in our timeseries observations.

\begin{deluxetable}{crccl}
\tablecaption{Ground-based Photometry of \thisstar}
\tablehead{\colhead{Planet} &
           \colhead{Facility} &
           \colhead{Filter} &
           \colhead{Type\tablenotemark{a}} &
           \colhead{$T_c$} \\
           \colhead{} &
           \colhead{} &
           \colhead{} &
           \colhead{} &
           \colhead{BTJD$_{\rm TDB}$\tablenotemark{b}}
           }
\startdata
b & TRAPPIST-S $0.6$-m\tablenotemark{c} 
                    & z$^\prime$ & F & 1378.6245 \\ 
b & LCO-SS     $0.4$-m & i$^\prime$ & I & 1383.2783 \\ 
b & SLR2       $0.5$-m & V          & E & 1383.2783 \\ 
b & SLR2       $0.5$-m & V          & I & 1392.5860 \\ 
b & LCO-SAAO   $1.0$-m & z$_{\rm s}$& I & 1392.5860 \\ 
b & LCO-SAAO   $1.0$-m & i$^\prime$ & I & 1406.5475 \\ 
c & MKO CDK700 $0.7$-m & r$^\prime$ & I & 1371.0589 \\ 
c & LCO-SAAO   $1.0$-m & z$_{\rm s}$& F & 1389.3608 \\ 
c & IRSF       $1.4$-m & J, H, K$_{\rm s}$& F & 1398.5117 \\ 
c & SSO/Europa $1.0$-m\tablenotemark{c} 
                       & z$^\prime$ & F & 1407.6626 \\ 
.03 & LCO-CTIO $1.0$-m & i$^\prime$ & I & 1442.7549 \\
\enddata
  \tablenotetext{a}{
    \footnotesize{F: full transit (covering  ingress and egress); I: ingress only; E: egress only}
  }
  \tablenotetext{b}{
    \footnotesize{Times of conjunction are given in the standard \tess-reported format, which is BJD$_{\rm TDB}-2457000$.}
  }
  \tablenotetext{c}{
    \footnotesize{TRAPPIST \citep{trappist}; SPECULOOS \citep{speculoos}}
  }  
  \tablecomments{
    \footnotesize{Each ground-based follow-up light curve is listed here, along with the predicted time of transit. Because the transits are so shallow ($<1$\ part per thousand for all three candidates), the ground-based data do not confidently detect the transits, and we do not include them in the global fit (\rfsecl{exofast}). No nearby EBs were detected.}
  }
\label{tab:sg1_table}
\end{deluxetable}

\vspace{20pt}

\subsection{Archival Imaging}
\label{sec:patient}

Ground-based photometric follow-up can rule out the presence of EBs at modest separations, but a physically unassociated background star within a few arcseconds of the location of \thisstar\ could plausibly produce a transit-like signal that
would not be resolved as a separate source in the few-arcsecond PSFs of follow-up images. To address this possibility, we examine archival images, in which the proper motion of \thisstar\ has carried it away from its current location. \rffigl{patient}\ shows the \tess\ image from Sector 1 along with images from the
ESO/SERC Southern Sky Atlas (SERC-J; taken in 1975) and the Anglo-Australian Observatory Second Epoch Survey (AAO-SES; 1993). The most constraining of these is the SERC-J image, enlarged in the left panel. The proper motion of $\mu_{\alpha} = -120$\,\masyr\ and $\mu_{\delta} = -123$\,\masyr\ leads to motion
of $1\farcs7$\ per decade; in the $43$\ years since the SERC-J image was obtained, \thisstar\ has moved $7\farcs4$. A background source at the current location of \thisstar should be seen as elongation of the PSF or a nearly resolved source. There is no indication of such features in either the blue-sensitive SERC-J or the red-sensitive AAO-SES images, so we conclude that there is no background source coincident with the present-day location of \thisstar.

\subsection{High Angular Resolution Imaging}
\label{sec:speckle}

Ground-based photometry rules out EBs at modest separations and archival imaging rules out background sources, but there may still be a bound stellar companion at small angular separation. An unresolved companion may itself be an EB responsible for one of the transit-like signals, but even if it is not, the dilution must be taken into account in the light curve fit in order to derive accurate radii \citep[e.g.,][]{buchhave:2011}, and the presence (or absence) of a binary companion can help us understand the formation of compact planetary systems \cite[e.g.,][]{kraus:2016}. Fortunately, bound companions to \tess\ planet hosts will be more easily revealed by high-resolution imaging than the typical \kep\ system because they are, on average, more nearby \citep[e.g.,][]{ciardi:2015,matson:2019}.

\begin{figure}
    \centering
    \includegraphics[width=\columnwidth]{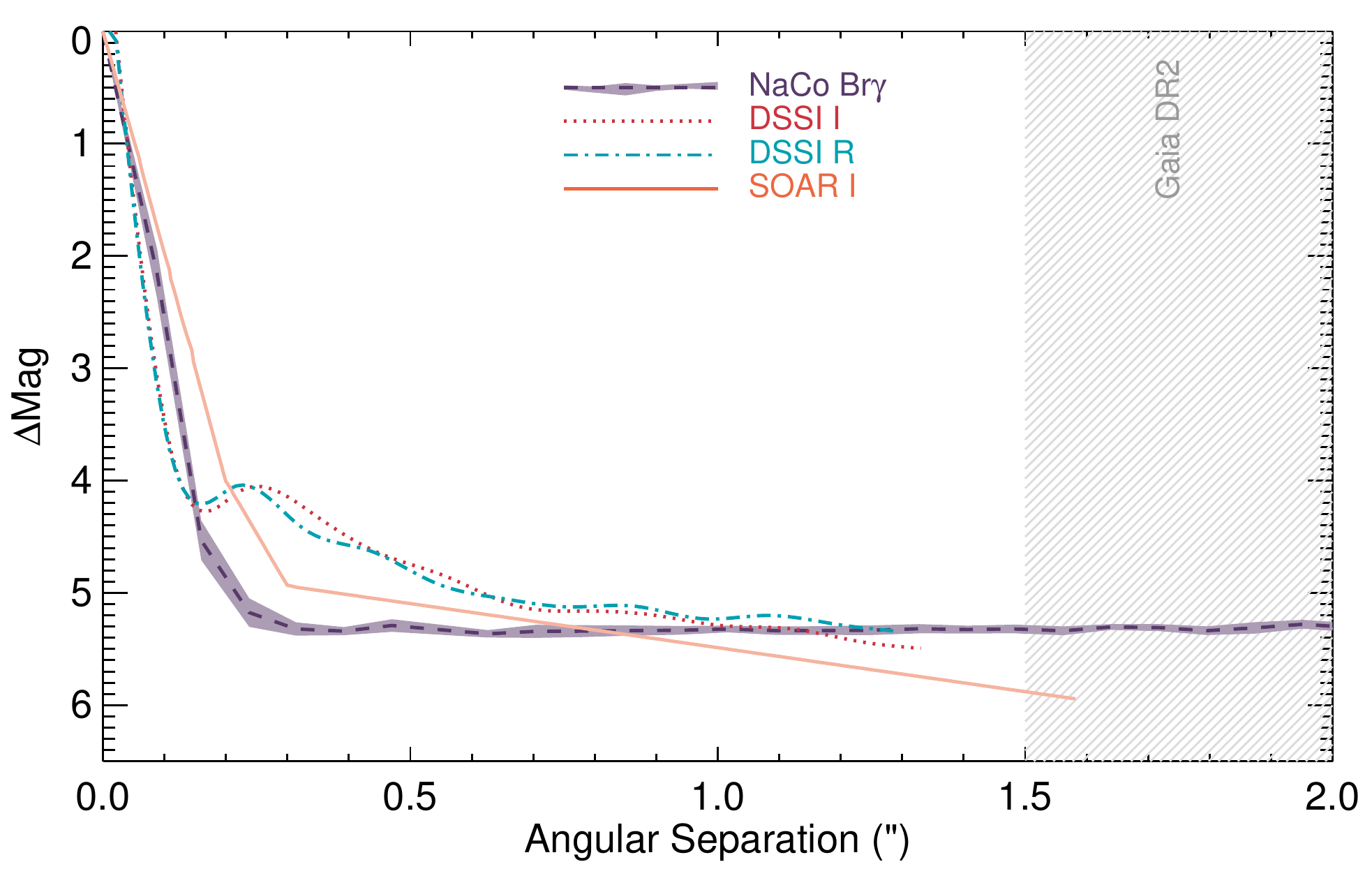}
    \caption{We show the $5$-$\sigma$ contrast curves for the high resolution imaging observations of \thisstar: SOAR HRCam speckle imaging in $I$-band (solid orange line); Gemini DSSI speckle imaging in $R$-band (blue dot-dashed line) and $I$-band (red dotted line); and VLT NaCO AO imaging in $Br\gamma$ (purple dashed line, with azimuthal scatter shown as a light purple shaded region). We exclude companions fainter by up to about 5 magnitudes in all bands outside a few tenths of an arcsecond. Gaia DR2 excludes the presence of wider companions bright enough to produce the $\mysim1$\ mmag signals of the high-SNR candidates.}
    \label{fig:speckle}
\end{figure}

We searched for close companions
to \thisstar\ in $I$-band using the HRCam speckle imager on the $4.1$-m Southern Astrophysical Research (SOAR) telescope \citep{Tokovinin:2018, Ziegler:2018a} on 2018 September 25 UT, in narrow-band $Br\gamma$ using the NaCo adaptive optics imager \citep{rousset:1998,lenzen:1998} on the 8-m UT1 of the VLT on 2018 October 23, and simultaneously in $R$- and $I$-band using the DSSI speckle imager \citep{horch:2009,horch:2012} on the 8-m Gemini South Telescope on 2018 October 31 UT. We detected no companions in any of these images down to contrast ratios of more than $5$\ magnitudes outside of $0\farcs2$ of \thisstar. Outside of $1\farcs5$, Gaia DR2 can exclude the presence of stellar sources bright enough to cause the $\mysim 1$\ mmag transit signals when blended with \thisstar\ \citep{ziegler:2018b}. The $5$-$\sigma$ contrast curves are shown in \rffigl{speckle}. 

\subsection{Reconnaissance Spectroscopy} 
\label{sec:recon}

We obtained three spectra with the CORALIE spectrograph \citep{queloz:2000,pepe:2017} on the Swiss Euler 1.2-m telescope of the ESO-La Silla Observatory (Chile) between UT 2018 Sept 07 and 2018 Oct 02. We used simultaneous Fabry-P\'{e}rot calibration for intrinsic drift measurement. The SNR per pixel of the individual spectra was $\mysim 20$.
Data were reduced using an adapted version of the HARPS pipeline: the average stellar line profiles, or cross-correlation functions (CCFs), were computed by cross-correlating the CORALIE spectra with a weighted binary G2 mask from which various tellurics and ISM lines were removed \citep{pepe:2002}. We see no evidence for multiple peaks in the CCF, suggesting that \thisstar\ does not have a bright, unresolved stellar companion. The RVs, reported in \rftabl{rv_table}, show no significant velocity variation. We derive spectroscopic parameters using SpecMatch \citep{petigura:2015,yee:2017}, and find $\teff=5187\pm110$\,K, $\logg=4.52\pm0.12$, $\feh=0.06\pm0.09$, and $\vsini<2$\,\kms. We use these values as starting guesses for our global model, and apply the derived \feh\ as a prior (see \rfsecl{exofast}).

\begin{deluxetable}{rrrrr}
\tablecaption{CORALIE radial velocities of \thisstar}
\tablehead{\colhead{${\rm BJD_{TDB}}$} &
           \colhead{RV} &
           \colhead{$\sigma_{\rm RV}$} &
           \colhead{BIS} &
           \colhead{$\sigma_{\rm BIS}$} \\
           \colhead{} &
           \colhead{\kms} &
           \colhead{\kms} &
           \colhead{\kms} &
           \colhead{\kms}
           }
\startdata
$2458368.687418$ & $11.071$ & $0.011$ & $-0.082$ & $0.011$ \\
$2458379.908910$ & $11.047$ & $0.018$ & $-0.077$ & $0.018$ \\
$2458393.713423$ & $11.064$ & $0.013$ & $-0.073$ & $0.013$ \\
\enddata
  \tablecomments{
    \footnotesize{Radial velocities and bisector span measurements from CORALIE observations (\rfsecl{recon}). We detect no variation in either quantity, consistent with expectations for a quiet star orbited by small planets.}
  }
\label{tab:rv_table}
\end{deluxetable}

\begin{table}
\scriptsize
\centering
\caption{Literature Properties for \thisstar}
\begin{tabular}{llcc}
  \hline
  \hline
 & \multicolumn{3}{c}{TIC 52368076} \\
Other & \multicolumn{3}{c}{TYC 88956-00192-1} \\
identifiers	  & \multicolumn{3}{c}{2MASS J01342273-6640328} \\
 & \multicolumn{3}{c}{Gaia DR2 4698692744355471616} \\
\hline
\hline
Parameter & Description & Value & Source\\
\hline 
$\alpha_{J2000}$\dotfill	&Right Ascension (RA)\dotfill & 01:34:22.735& 1	\\
$\delta_{J2000}$\dotfill	&Declination (Dec)\dotfill & -66:40:32.95& 1	\\
\\
$T$\dotfill	& \tess\ $T$ mag.\dotfill	& $10.138\pm0.017$ & 6	\\
\\
$B_T$\dotfill	& Tycho B$_T$ mag.\dotfill & 11.882 $\pm$ 0.077		& 2	\\
$V_T$\dotfill	& Tycho V$_T$ mag.\dotfill & 11.102 $\pm$ 0.065		& 2	\\
$B$\tablenotemark{a}\dotfill		
                & APASS Johnson $B$ mag.\dotfill	& 11.701 $\pm$	0.025& 3	\\
$V$\dotfill		& APASS Johnson $V$ mag.\dotfill	& 10.892 $\pm$	0.016& 3	\\
$G$\dotfill     & Gaia $G$ mag.\dotfill     &   10.7180 $\pm$ 0.0004    & 1\\
$g'$\dotfill	& APASS Sloan $g'$ mag.\dotfill	& 11.268$\pm$0.019	& 3	\\
$r'$\dotfill	& APASS Sloan $r'$ mag.\dotfill	& 10.458$\pm$0.041	& 3	\\
$i'$\dotfill	& APASS Sloan $i'$ mag.\dotfill	& 10.662$\pm$0.017 & 3	\\
\\
$J$\dotfill		& 2MASS $J$ mag.\dotfill & 9.466  $\pm$ 0.02	& 4	\\
$H$\dotfill		& 2MASS $H$ mag.\dotfill & 9.112 $\pm$ 0.03	    & 4	\\
$K_S$\dotfill	& 2MASS $K_S$ mag.\dotfill & 8.995 $\pm$ 0.02& 4	\\
\\
\textit{WISE1}\dotfill		& \textit{WISE1} mag.\dotfill & 8.945 $\pm$ 0.03 	& 5	\\
\textit{WISE2}\dotfill		& \textit{WISE2} mag.\dotfill & 9.006 $\pm$ 0.03 	& 5 \\
\textit{WISE3}\dotfill		& \textit{WISE3} mag.\dotfill & 8.944 $\pm$ 0.03 	& 5	\\
\textit{WISE4}\dotfill		& \textit{WISE4} mag.\dotfill & 8.613 $\pm$0.262 	& 5	\\
\\
$\mu_{\alpha}$\dotfill		& PM in RA (mas yr$^{-1}$) \dotfill & -119.800 $\pm$ 0.066	& 1 \\
$\mu_{\delta}$\dotfill		& PM in DEC (mas yr$^{-1}$) \dotfill  	&  -122.953 $\pm$ 0.080 &  1 \\
$\pi$\dotfill & Parallax (mas) \dotfill & 8.976 $\pm$ 0.036 & 1 \\
$RV$\dotfill & Systemic RV (\kms) \dotfill  & 11.062 $\pm$ 0.012 & 7 \\
\hline
\\[-6ex]
\end{tabular}
\begin{flushleft} 
\tablenotetext{a}{
    \footnotesize{The uncertainties of the photometry have a systematic error floor applied. Even still, the global fit requires a significant scaling of the uncertainties quoted here to be consistent with our model, suggesting they are still significantly underestimated for one or more of the broad band magnitudes.
    }
}
\tablecomments{
    \footnotesize{References are:
    $^1$\citet{Gaia:2018};
    $^2$\citet{Hog:2000};
    $^3$\citet{Henden:2016};
    $^4$\citet{Cutri:2003};
    $^5$\citet{Cutri:2014};
    $^6$\citet{stassun:2018b};
    $^7$This work
    }
}
\end{flushleft}
\label{tab:LitProps}
\end{table}

\clearpage

\section{EXOFASTv2 Global Fit} 
\label{sec:exofast}

\begin{figure}[!t]
\centering 
\includegraphics[width=\columnwidth]{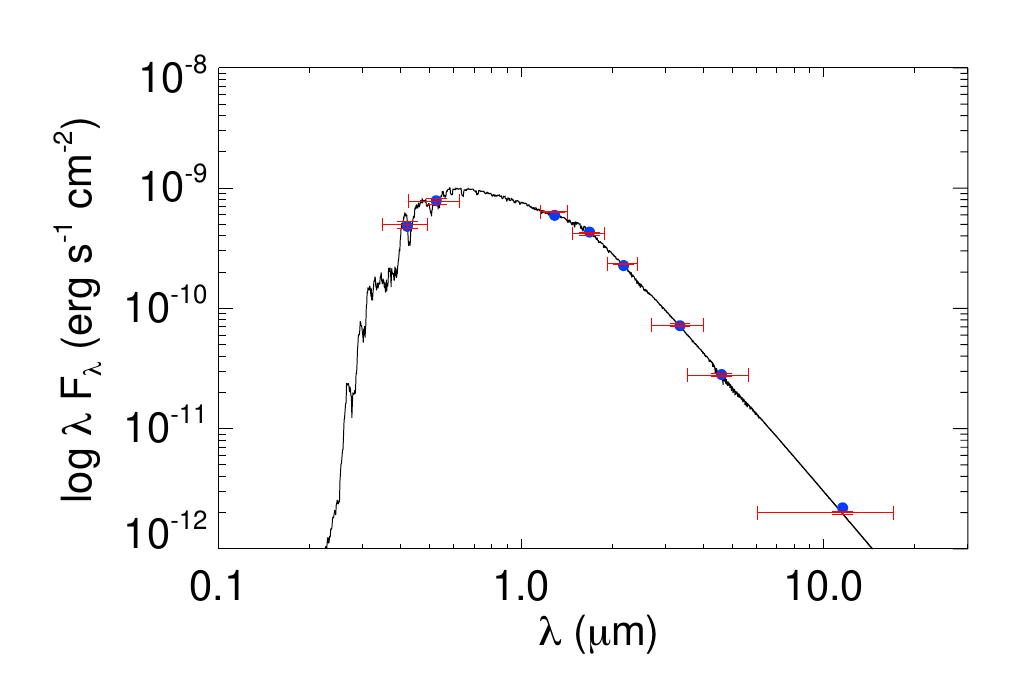}
\caption{The spectral energy distribution fit from EXOFASTv2 for \thisstar. The red points are the observed values at the corresponding passbands and the blue points are the predicted integrated fluxes. The horizontal red error bars represent the width of the bandpasses and the vertical errors represent the 1$\sigma$ uncertainties. The final model fit is shown as a solid black line. }
\label{fig:sed_fit}
\end{figure}

To gain a full understanding of the system parameters, we globally fit the available photometric and spectroscopic data using the publicly available exoplanet modeling suite, EXOFASTv2 \citep{Eastman:2013, Eastman:2017}. Specifically, we fit the \tess\ light curves from observing Sectors 1 and 2 for planets b, c, and candidates .03, .04, and .05 (See Figures \ref{fig:lc}, \ref{fig:usp}, and \ref{fig:cand}), while constraining the host star parameters using the spectral energy distribution (SED) and the MESA Isochrones and Stellar Tracks (MIST) stellar isochrones \citep{Dotter:2016, Choi:2016, Paxton:2011, Paxton:2013, Paxton:2015}. The broadband photometry is given in \rftabl{LitProps} and shown along with the final model in \rffigl{sed_fit}. We enforce Gaussian priors $\teff = 5187\pm110$\ K and $\feh = 0.06\pm0.09$\ dex from the analysis of the CORALIE spectra. We also place a conservative Gaussian prior on the parallax from {\it Gaia} DR2 of $8.976\pm0.1$\ mas, since all possible uncertainties should total to less than 0.1 mas \citep{Gaia:2016, Gaia:2018}. Lastly, we enforce an upper limit on the extinction of A$_V$ = 0.0521 from the Schlegel Galactic dust reddening and extinction maps \citep{Schlegel:1998}. All other parameters were allowed to vary without prior constraints. We allowed an error scaling term for the SED photometry (reported in \rftabl{exofast_stellar}) and a variance term for each sector of the TESS photometry (\rftabl{exofast}). Limb darkening parameters are interpolated using the current \logg, \teff, and \feh\ at each step and the limb-darkening tables of \citet{Claret:2017}. We adopt the strict convergence criteria recommended by \citet{Ford:2006} in order to assure the global minimum has been identified and covariances are well characterized: the Gelman-Rubin statistic for all parameters must be less than $1.01$, and the number of independent draws (chain length divided by correlation length) must exceed $1000$. We ran a fit that allowed transit timing variations (TTVs) but find no significant TTVs and no changes to the derived parameters, so for simplicity we adopt the solution that assumes periodic ephemerides. We also ran fits including only the two, three, or four strongest signals, and we find stellar and planetary results fully consistent with the five planet solution. Because inclusion of the two more marginal candidates does not affect our conclusions about \thisstar b, c, and .03, we present herein the 5-planet fit. The final system parameters determined by the EXOFASTv2 fit, including predicted masses using the relations of \citep{Chen:2017}, are shown in \rftabs{exofast_stellar}{exofast}. We again refer the reader to the top view of the system architecture in \rffigl{arch}.

\begin{table}
\small
\centering
\caption{\thisstar\ stellar parameters: median values and 68\% CI}
\begin{tabular}{llcccc}
  \hline
  \hline
Parameter & Units & Values & & & \\
\hline
\multicolumn{2}{l}{Stellar Parameters:}&\smallskip\\
~~~~$M_*$\dotfill &Mass (\msun)\dotfill &$0.871^{+0.046}_{-0.040}$\\
~~~~$R_*$\dotfill &Radius (\rsun)\dotfill &$0.852^{+0.017}_{-0.016}$\\
~~~~$L_*$\dotfill &Luminosity (\lsun)\dotfill &$0.509^{+0.024}_{-0.025}$\\
~~~~$\rho_*$\dotfill &Density (cgs)\dotfill &$1.99\pm0.15$\\
~~~~$\log{g}$\dotfill &Surface gravity (cgs)\dotfill &$4.518\pm0.027$\\
~~~~$T_{\rm eff}$\dotfill &Effective Temperature (K)\dotfill &$5282^{+67}_{-75}$\\
~~~~$[{\rm Fe/H}]$\dotfill &Metallicity (dex)\dotfill &$0.069^{+0.083}_{-0.081}$\\
~~~~$Age$\dotfill &Age (Gyr)\dotfill &$6.6^{+4.6}_{-4.2}$\\
~~~~$EEP$\dotfill &Equal Evolutionary Point \dotfill &$348^{+28}_{-27}$\\
~~~~$A_V$\dotfill &V-band extinction (mag)\dotfill &$0.024^{+0.019}_{-0.017}$\\
~~~~$\sigma_{SED}$\dotfill &SED photometry error scaling \dotfill &$2.45^{+1.1}_{-0.62}$\\
~~~~$\pi$\dotfill &Parallax (mas)\dotfill &$8.975^{+0.099}_{-0.100}$\\
~~~~$d$\dotfill &Distance (pc)\dotfill &$111.4^{+1.3}_{-1.2}$\\
\hline
\end{tabular}
\label{tab:exofast_stellar}
\end{table}

\begin{table*}
\scriptsize
\setlength{\tabcolsep}{2pt}
\centering
\caption{\thisstar\ planetary and transit parameters: median values and 68\% confidence interval}
\begin{tabular}{llccccc}
  \hline
  \hline
Parameter & Description (Units) & \multicolumn{5}{c}{Values} \\
\hline
 & & Low SNR & {\bf Validated} & {\bf Validated} & Marginal & {\bf High SNR} \\
 & & .04 & b & c & .05\tablenotemark{a} & .03 \\
~~~~$P$\dotfill &Period (days)\dotfill &$0.528474^{+0.000040}_{-0.000030}$&$4.65382^{+0.00032}_{-0.00031}$&$9.15067^{+0.00062}_{-0.00069}$&$13.2781^{+0.0020}_{-0.0019}$&$19.9807^{+0.0045}_{-0.0049}$\\
~~~~$R_P$\dotfill &Radius (\re)\dotfill &$1.36^{+0.14}_{-0.16}$&$2.755^{+0.091}_{-0.079}$&$2.79\pm0.10$&$8.8^{+4.7}_{-4.4}$&$2.94\pm0.16$\\
~~~~$T_C$\dotfill &Time of conjunction (\bjdtdb)\dotfill &$2458350.8394\pm0.0011$&$2458355.35520^{+0.00093}_{-0.00087}$&$2458352.7582^{+0.0014}_{-0.0013}$&$2458365.0560^{+0.0019}_{-0.0020}$&$2458342.8514^{+0.0034}_{-0.0033}$\\
~~~~$a$\dotfill &Semi-major axis (AU)\dotfill &$0.01222^{+0.00021}_{-0.00019}$&$0.05210^{+0.00090}_{-0.00082}$&$0.0818^{+0.0014}_{-0.0013}$&$0.1048^{+0.0018}_{-0.0016}$&$0.1376^{+0.0024}_{-0.0022}$\\
~~~~$i$\dotfill &Inclination (Degrees)\dotfill &$72.80^{+0.72}_{-0.70}$&$88.99^{+0.70}_{-0.81}$&$88.52^{+0.32}_{-0.19}$&$87.70^{+0.15}_{-0.14}$&$88.753^{+0.080}_{-0.081}$\\
~~~~$e$\dotfill &Eccentricity \dotfill &--&$0.183^{+0.14}_{-0.098}$&$0.065^{+0.067}_{-0.046}$&$0.037^{+0.046}_{-0.027}$&$0.075^{+0.056}_{-0.051}$\\
~~~~$\omega_*$\dotfill &Argument of Periastron (Degrees)\dotfill &--&$-91^{+57}_{-56}$&$90^{+97}_{-98}$&$50\pm120$&$90^{+100}_{-110}$\\
~~~~$ecos{\omega_*}$\dotfill & \dotfill &--&$-0.00\pm0.17$&$-0.000^{+0.055}_{-0.054}$&$0.000^{+0.036}_{-0.034}$&$0.000\pm0.063$\\
~~~~$esin{\omega_*}$\dotfill & \dotfill &--&$-0.114^{+0.057}_{-0.098}$&$0.014^{+0.075}_{-0.042}$&$0.001^{+0.037}_{-0.031}$&$0.011^{+0.072}_{-0.053}$\\
~~~~$\fave$\dotfill &Incident Flux (\fluxcgs)\dotfill &$4.63\pm0.25$&$0.243^{+0.017}_{-0.019}$&$0.1026^{+0.0056}_{-0.0055}$&$0.0627^{+0.0034}_{-0.0033}$&$0.0362^{+0.0020}_{-0.0019}$\\
~~~~$T_{eq}$\dotfill &Equilibrium temperature (K)\dotfill &$2126^{+28}_{-29}$&$1029\pm14$&$821\pm11$&$725.8^{+9.5}_{-9.8}$&$633.5^{+8.3}_{-8.5}$\\
~~~~$R_P/R_*$\dotfill &Radius of planet in stellar radii \dotfill &$0.0146^{+0.0014}_{-0.0016}$&$0.02962^{+0.00077}_{-0.00063}$&$0.02998^{+0.00087}_{-0.00086}$&$0.095^{+0.051}_{-0.047}$&$0.0317^{+0.0015}_{-0.0016}$\\
~~~~$a/R_*$\dotfill &Semi-major axis in stellar radii \dotfill &$3.085\pm0.077$&$13.15\pm0.33$&$20.65\pm0.52$&$26.47\pm0.66$&$34.75\pm0.87$\\
~~~~$d/R_*$\dotfill &Separation at mid transit \dotfill &$3.085\pm0.077$&$14.29^{+1.3}_{-0.97}$&$20.2^{+1.1}_{-1.5}$&$26.4^{+1.1}_{-1.2}$&$34.1^{+2.2}_{-2.5}$\\
~~~~$\delta$\dotfill &Transit depth (fraction)\dotfill &$0.000214^{+0.000043}_{-0.000045}$&$0.000877^{+0.000046}_{-0.000037}$&$0.000899^{+0.000053}_{-0.000051}$&$0.0090^{+0.012}_{-0.0067}$&$0.001004^{+0.00010}_{-0.000096}$\\
~~~~$\tau$\dotfill &Ingress/egress transit duration (days)\dotfill &$0.00206^{+0.00049}_{-0.00039}$&$0.00376^{+0.00095}_{-0.00024}$&$0.00488^{+0.00070}_{-0.00078}$&$0.0232^{+0.0018}_{-0.0015}$&$0.0086^{+0.0016}_{-0.0014}$\\
~~~~$T_{14}$\dotfill &Total transit duration (days)\dotfill &$0.0255^{+0.0022}_{-0.0027}$&$0.1233^{+0.0025}_{-0.0026}$&$0.1227^{+0.0025}_{-0.0028}$&$0.0464^{+0.0037}_{-0.0030}$&$0.1284^{+0.0055}_{-0.0057}$\\
~~~~$T_{FWHM}$\dotfill &FWHM transit duration (days)\dotfill &$0.0235^{+0.0025}_{-0.0031}$&$0.1192^{+0.0023}_{-0.0024}$&$0.1179^{+0.0026}_{-0.0030}$&$0.0232^{+0.0019}_{-0.0015}$&$0.1198^{+0.0060}_{-0.0063}$\\
~~~~$b$\dotfill &Transit Impact parameter \dotfill &$0.912^{+0.023}_{-0.022}$&$0.25^{+0.24}_{-0.17}$&$0.524^{+0.078}_{-0.14}$&$1.056^{+0.055}_{-0.057}$&$0.745^{+0.047}_{-0.060}$\\
~~~~$M_P$\dotfill &Predicted Mass (\me)\dotfill &$2.65^{+0.94}_{-0.56}$&$8.5^{+2.8}_{-1.8}$&$8.6^{+2.8}_{-1.9}$&$61^{+200}_{-42}$&$9.5^{+3.2}_{-2.1}$\\
~~~~$K$\dotfill &Predicted RV semi-amplitude (m/s)\dotfill &$2.19^{+0.78}_{-0.47}$&$3.65^{+1.2}_{-0.80}$&$2.88^{+0.95}_{-0.64}$&$18^{+59}_{-13}$&$2.45^{+0.84}_{-0.56}$\\
~~~~$logK$\dotfill &Log of RV semi-amplitude \dotfill &$0.34^{+0.13}_{-0.10}$&$0.56^{+0.12}_{-0.11}$&$0.46^{+0.12}_{-0.11}$&$1.26^{+0.63}_{-0.52}$&$0.39^{+0.13}_{-0.11}$\\
~~~~$\rho_P$\dotfill &Predicted Density (cgs)\dotfill &$5.9^{+2.0}_{-1.3}$&$2.21^{+0.72}_{-0.47}$&$2.17^{+0.70}_{-0.47}$&$0.69^{+0.92}_{-0.36}$&$2.04^{+0.68}_{-0.44}$\\
~~~~$logg_P$\dotfill &Predicted Surface gravity \dotfill &$3.147^{+0.12}_{-0.092}$&$3.04^{+0.12}_{-0.10}$&$3.03^{+0.12}_{-0.10}$&$2.94^{+0.26}_{-0.15}$&$3.03^{+0.12}_{-0.10}$\\
~~~~$M_P/M_*$\dotfill &Predicted Mass ratio \dotfill &$0.0000091^{+0.0000033}_{-0.0000020}$&$0.0000291^{+0.0000098}_{-0.0000064}$&$0.0000295^{+0.0000098}_{-0.0000066}$&$0.00021^{+0.00069}_{-0.00015}$&$0.0000325^{+0.000011}_{-0.0000075}$\\
[1ex]
\hline
\multicolumn{2}{l}{Wavelength Parameters:}&TESS\smallskip\\
~~~~$u_{1}$\dotfill &linear limb-darkening coeff \dotfill &$0.391\pm0.037$\\
~~~~$u_{2}$\dotfill &quadratic limb-darkening coeff \dotfill &$0.231\pm0.036$\\[1ex]
\hline
\multicolumn{2}{l}{Transit Parameters:}&TESS Sector 1 & TESS Sector 2\smallskip\\
~~~~$\sigma^{2}$\dotfill &Added Variance \dotfill &$4.0^{+1.8}_{-1.7}\times10^{-8}$&$1.1\pm1.7\times10^{-8}$\\
~~~~$F_0$\dotfill &Baseline flux \dotfill &$1.000033\pm0.000013$&$1.000074\pm0.000012$\\
\hline \\[-6ex]
\end{tabular}
  \begin{flushleft} 
  \tablenotetext{a}{\footnotesize{The values reported for \thisstarfive\ are those from EXOFASTv2 without any radial velocity constraint. We note that when we exclude those solutions inconsistent with the CORALIE RVs, the most probable size is $R_P\mysim 4.2$\ \rearth (see \rffig{cand}). Moreover, planets like \thisstarfive\ are a priori likely to be small; if real, its true size is probably even smaller.}}
  \tablecomments{\footnotesize{We reiterate that \thisstarb, c, and .03 are high-SNR events; the USP \thisstarfour\ is intriguing even though it does not meet our formal significance threshold; and we consider \thisstarfive\ a marginal planet candidate but present it for completeness.}
  }
  \end{flushleft}
\label{tab:exofast}
\end{table*}

\section{Statistical Validation with VESPA}
\label{sec:vespa}

We used the \vespa\ package \citep{morton:2015} to assess the statistical likelihood that the transits of \thisstarb\ and c are caused by planets, rather than false positives. \vespa\ simulates stellar and planetary systems to generate transits (and eclipses) to compare against the observed data of \thisstar. Rejecting systems that are inconsistent with the observations, \vespa\ then calculates the false positive probability (FPP) for each candidate. The FPP depends not only on the transit shape, but also the position of the star on the sky (to assess the likelihood of background blends), the stellar parameters (which inform not only transit and eclipse shapes but also the likelihood of stellar companions), the extent to which nearby EBs can be excluded by high resolution imaging, and the presence or absence of features in the light curve that might indicate the presence of a binary (such as depth differences in alternating transits or the presence of a secondary eclipse). We therefore provide to \vespa\ the sky coordinates, stellar parameters, literature photometry, high resolution imaging contrast curves, and the flattened \tess\ light curve. Our RVs rule out an EB (as opposed to a blended background or hierarchical EB).

After running \vespa, we adjusted the FPP by excluding the scenario in which a direct EB companion to \thisstar\ causes one of the transit signals. The resulting FPPs are $6\times10^{-5}$\ and $9\times10^{-5}$\ for \thisstarone\ and .02, respectively. Note that we did not remove the contribution from background EBs even though our inspection of the archival imaging suggests there are no background stars at the current location of \thisstar; the FPPs would be even lower with this adjustment. We therefore conclude that these are statistically validated planets, and now refer to them as \thisstarb\ and c. We do not attempt to validate \thisstarthree\ despite its high SNR, because we observed only two transits. Similarly, we do not attempt to validate \thisstarfour\ and .05 because \vespa\ does not assess the likelihood that a signal is an instrumental false alarm, and we cannot fully exclude this possibility given the low SNR of these events.

\section{Dynamics}
\label{sec:dynamics}
This section considers dynamical aspects of the \thisstar\ system, considering the three planets with the highest S/N detections.
We first note that \thisstarb\ has non-zero eccentricity, but a relatively short time scale for the damping of its eccentricity. 
As a result, the system is dynamically interesting, suggesting some type of planet-planet interactions. These types of interactions could lead to TTVs (\rfsecl{ttvs}), although they are not observed in the present data set. We also need to consider possible instability of the system (\rfsecl{stability}), which puts additional constraints on the allowed current-day values of the orbital eccentricities. 
Finally, we consider the question of mean motion resonances, and find that the system is highly unlikely to currently be in a true resonant configuration (\rfsecl{mmr}).

\subsection{Transit Timing Variations (TTVs)} 
\label{sec:ttvs}
The proximity of \thisstar b and c to a 2:1 MMR means that their mutual gravitational perturbations add in a nearly-coherent manner that can lead to significant and potentially-measurable TTVs \citep[e.g.,][]{Agol:2005,Holman:2005}. 
An attempt to model the planets' timing variations in the first two sectors of \tess\ did not yield significant dynamical constraints, as the uncertainty on the transit times from \tess\ are larger than the expected TTVs. Here we briefly consider the prospects for extracting dynamical information by combining the first two sectors of \tess\ observations with future follow-up observations. 

\begin{figure}[!t]
    \centering
    \includegraphics[width=\columnwidth]{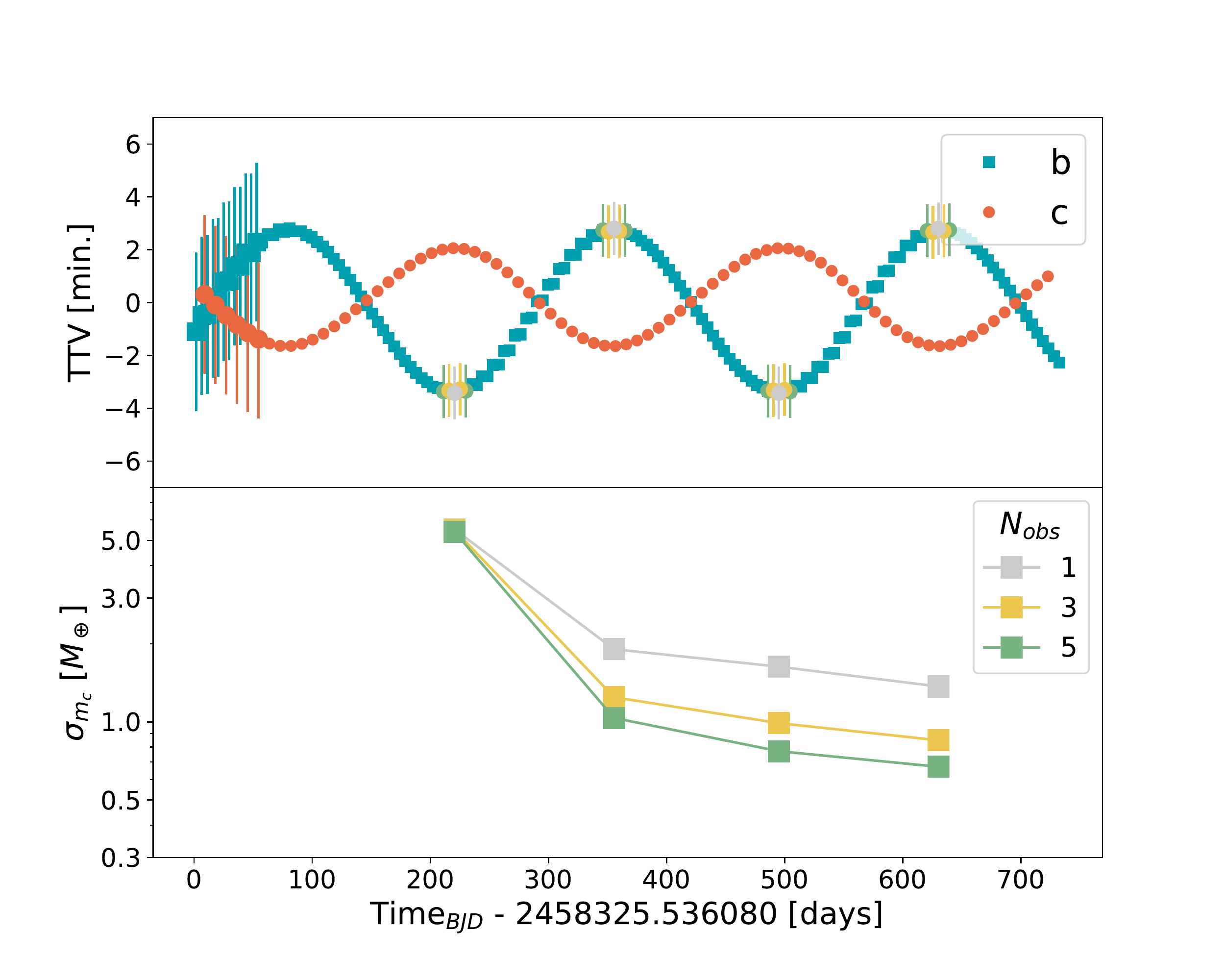}
    \caption{The top panel shows predicted transit times of \thisstar b (blue) and c (orange) with representative $1$-$\sigma$ error bars from the first two TESS sectors. The projected TTV signals assume fiducial planet masses of $8.6\mearth$ and a stellar mass $M_*=0.87\msun$. A series of hypothetical follow-up transit observations of planet b with one-minute transit mid-time uncertainties are shown by colored points at the peaks of planet b's TTV signal. Different color points correspond to follow-up scenarios in which 1, 3 or 5 transit observations are obtained at each epoch. The bottom panel shows the expected $1$-$\sigma$ uncertainty in planet c's mass, $\sigma_c$, that would be achieved by these follow-up transit observations. }
    \label{fig:ttv}
\end{figure}

To assess the possibility for extracting dynamical information from \thisstar b and c's TTVs, we need to know how the expected TTV signal amplitude compares to the precision of any future transit time measurements. We employ the analytic TTV model detailed in \cite{Hadden2018TTV} in order to predict the planets' TTV signals and the accompanying dynamical constraints derived from them. Figure \ref{fig:ttv} shows the TTV signals predicted for b and c, assuming no free eccentricity for either planet\footnote{
    Due to its short orbital period, the eccentricity damping timescale of planet b is short.
    Adopting the best-fit stellar mass, planet radius, and orbital period from \rftabl{exofast}, the tidal eccentricity damping timescale for \thisstar b is given by 
    \begin{eqnarray}\label{eq:circ}
        \left(\frac{1}{e_b}\frac{de_b}{dt}\right)^{-1} = 83~\text{Myr}\times\fracbrac{Q/k_2}{10^{3}}\fracbrac{m_b}{10\mearth}.
    \end{eqnarray}
    where $Q$ is the planet's tidal quality factor and $k_2$ its tidal Love number \citep{Goldreich1966}. The parenthetical terms on the right hand side are of order unity. With the same assumptions about $Q/ k_2$ for planet c, its nominal eccentricity damping timescale is $1.9~\text{Gyr}$. Dynamical coupling between b and c should enhance the efficiency with which planet c's eccentricity is damped.
    } 
and fiducial masses of $3\times10^{-5}\,M_*$ for both planets. The resulting TTVs are approximately sinusoidal with amplitudes of $\mysim 3$\ minutes. The planets' TTV signals will, to excellent approximation, simply scale linearly with the planet masses.
To estimate the timing precision one could achieve using {\it Spitzer}, we scale SNR from existing transit observations and employ the analytic formulae of \citet{carter:2008} and \citet{price:2014}, which lead to a per-transit uncertainty of $1$\ min.
    The bottom panel of Figure \ref{fig:ttv} shows the estimated precision with which planet c's mass could be measured using planet b's TTV by obtaining a series of follow-up observations centered on the peaks of the approximately sinusoidal TTV and assuming transit mid-times are measured with $1\sigma$ uncertainties of $1$ minute. Precisions approaching $\sim 1M_\earth$ are achievable with a series of transit observations consisting of $\sim 3-5$ transit timing measurements made at 2 or 3 successive peaks of the TTV signal.
    Note, though, that these mass measurement precisions are based on a TTV model that assumes zero free eccentricity for \thisstar b and c; relaxing this assumption would significantly increase the mass uncertainty due to the mass-eccentricity degeneracy inherent to TTVs of planets near MMRs \citep[e.g.,][]{Lithwick2012}. On the other hand, the combination of TTVs and RVs would provide the strongest possible mass and eccentricity constraints, and in \rfsecl{rv} we argue that the planets should be amenable to RV follow-up.

\subsection{Dynamical Stability}
\label{sec:stability}

\begin{figure*}[t!]
\includegraphics[width=6.6in]{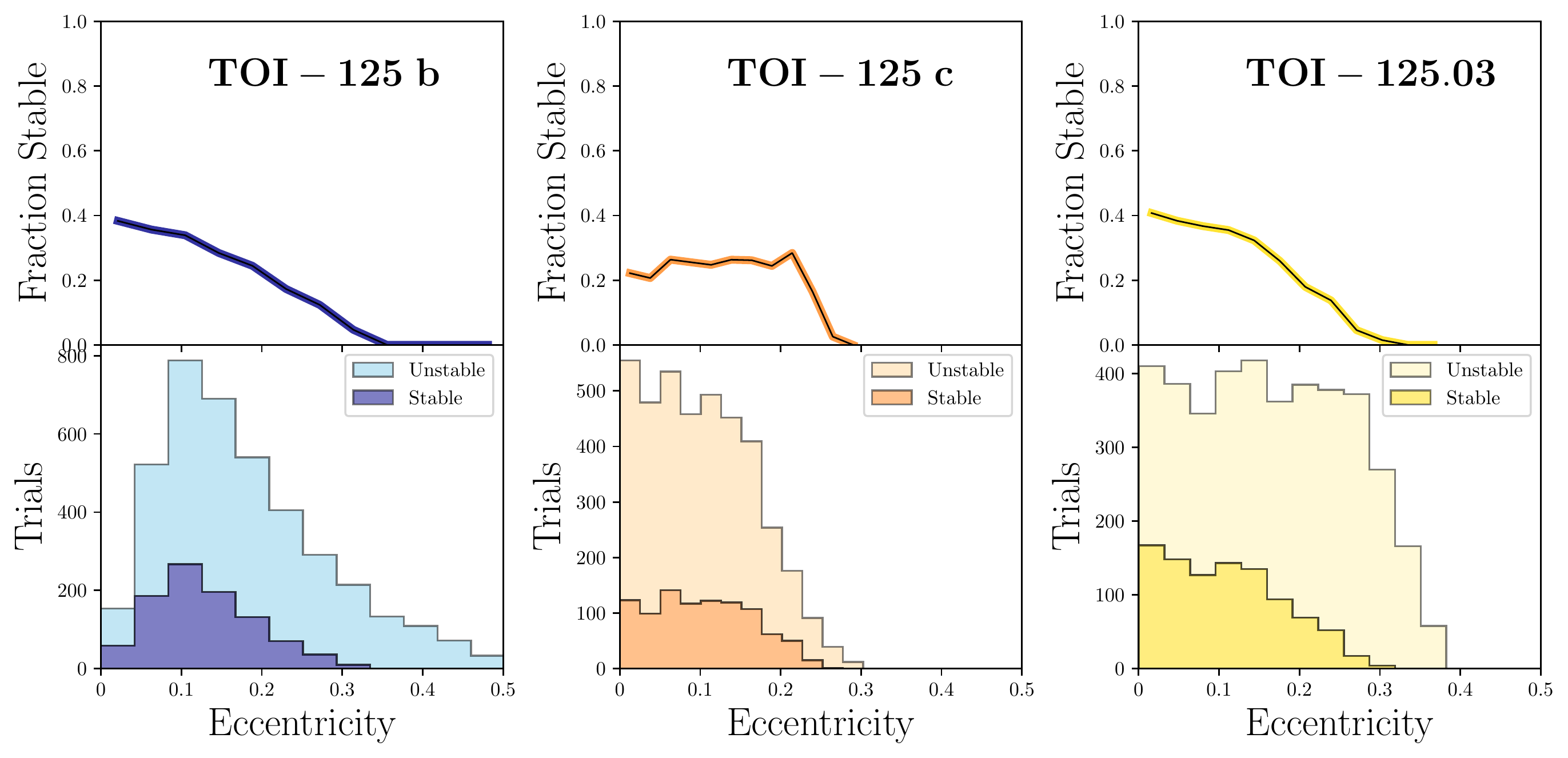}
\caption{For each planet, the dynamically allowed range of orbital eccentricities, derived from the suite of 1 Myr numerical simulations. (Top panels) For each planet,  the fraction of integrations in each planetary eccentricity bin that remains dynamically stable for the entire 1 Myr integration. (Bottom panels) For each planet, stacked histograms showing the stable and unstable trials for each eccentricity. The overall shape of the histograms is determined by the EXOFASTv2 fit results, from which initial simulation parameters were drawn.}
\label{fig:ecc_stable}
\end{figure*}

In the absence of external perturbations, a planet with an eccentric orbit residing close to its host star would generally become tidally circularized on astronomically short timescales (the exact timescale depending on the tidal quality factor of the planet and properties of its host star; see, e.g., \rfeql{circ}). 
Therefore, significant eccentricities for compact systems often require that the planets be located in regions of resonance \citep{Beauge:2006} and can result in significant transfer of angular momentum between planets \citep{Kane2014,Antoniadou2016}. 
Thus, a dynamical analysis of a proposed orbital solution can be used to validate or potentially revise the allowed architecture for the system. The EXOFASTv2 global model of the \thisstar multi-planet system cannot exclude a moderately large eccentricity of \thisstarb\ ($\sim0.18$), which would be unusual for a planet in a tightly-packed, multi-planet system \citep{Kane2012, Hadden:2014, VanEylen2015, Xie2016}. We therefore set out to investigate whether such high eccentricities can be ruled out through dynamical simulations, or if there exists evidence that \thisstarb\ and \thisstarc\ may be in resonance. 

In the analysis that follows, we consider only the three high-SNR transit signals (\thisstar b, c, and .03). The candidate USP planet (\thisstarfour), given its small size and orbital period, is effectively dynamically decoupled from the other planets and if it exists should not affect our conclusions. \thisstarfive\ {\it would} have to be incorporated in a dynamical investigation if it is shown to be real, but given its low SNR, we do not include it in our analysis for now. If follow-up observations show that \thisstarfive\ is real, it would reside slightly interior to the 3:2 mean motion commensurability (with a period ratio of $\sim1.45$), close to the regime in which resonant interactions are most relevant. 

To evaluate the dynamical stability and orbital evolution of the three planets in the \thisstar system, we performed 3000 numerical simulations using N-body code \texttt{Mercury6} \citep{Chambers:1999}, altered to include the effect of general relativistic precession. The simulations were performed using a hybrid symplectic and Bulirsch-Stoer (B-S) integrator with a time-step of 30 minutes for a total integration time of slightly more than 1 million years per integration (which is roughly 80 million orbits of the innermost planet), and energy was conserved to better than one part in $10^{9}$ (for energy changes due to the integrator). 
For each integration, we drew one link from the EXOFASTv2 transit fit posterior, and assigned planet and star properties equal to those in the chosen posterior link. 

These numerical simulations allow us to impose an additional level of constraints beyond those derived from the transit shapes: some planetary eccentricities will lead to dynamical instabilities in the system (which occur when scattering events or orbit crossing leads to physical collisions between planets, the ejection of a planet from the system, or collision of a planet with the central star). 
The composite eccentricity distributions (stacked stable + unstable) in Figure \ref{fig:ecc_stable} show the eccentricity draws for our 3000 simulations. The stable subset of each distribution contains those which allow the planets to remain dynamically stable for the entire 1 Myr integration. Not shown is the variation in other orbital elements (also drawn from the EXOFASTv2 posteriors), but the overall trend in stability fraction is shown in the top panels. 

Of these 3000 simulations, 32\% remained dynamically stable for the entire 1 Myr integration, and Figure \ref{fig:ecc_stable} show a higher stability fraction at lower values of eccentricity. Dynamical stability considerations thus prefer the lower values of eccentricity, with eccentricities above 0.25--0.3 disallowed for each planet. The conclusion from this analysis is that although the EXOFASTv2 posteriors allow an unusually large range of eccentricities, the true values are likely on the lower ends of these ranges. 

\subsection{Resonant Behavior and Formation History}
\label{sec:mmr}

The orbital periods of the two planets b and c (4.65437 and 9.1536 days) lie close to the mutual 2:1 commensurability. As a result, it is natural to wonder whether the two planets are trapped in mutual 2:1 mean motion resonance (with a period ratio of 1.967). 
As orbital elements (including semi-major axis) librate while planets reside in resonance, it is possible to reside in resonance even without a perfect 2:1 period ratio \citep[i.e.,]{Batygin:2013}.

However, the majority of planets near, but not in, orbital resonance reside slightly outside of a resonant configuration. The results of \citet[][see Figure 9]{Terquem2018} show that for planets migrating in a disk, it is very easy for the 2:1 resonance to be disrupted when the inner planet enters a disk cavity, as might occur at small orbital radii. The subsequent evolution of the system is more difficult to model, but it is expected that departures from resonance will move towards the outside of resonance in cases where
\begin{equation}
\delta = \frac{P_2}{P_1} - \frac{(q + 1)}{q} > -0.04.
\label{eq:resdelta}
\end{equation}
when the resonance considered is the $q + 1:q$ first order resonance. Since the inner two planets in the \thisstar system have $\delta$ = -0.033, they would be expected to fit this trend and reside slightly outside of the 2:1 MMR period ratio; however, these planets instead appear to have a period ratio slightly less than this value. There are three potential explanations for this:

{\it 1) These planets are currently in true orbital resonance}.
A system with a similar architecture to \thisstar is the Gliese-876 system \citep{Marcy:2001,Rivera2010}, which also has two planets close to the 2:1 resonance (Gliese-876 c and b, at $\sim30$ and $\sim60$ days), the inner of which exhibits significant (0.2) eccentricity. In the Gliese-876 system, those two planets form a Laplace resonance with a third planet. The non-zero planetary eccentricities are maintained through the resonance. The \thisstar system also has three high-SNR planets with orbital periods moderately close to 2:1 resonances. 
An orbital resonance which persisted after the disk dissipated could explain the large eccentricity of the inner planet. While the disk is present, the orbital eccentricity of both planets in the system will be damped. Once the disk has dissipated, the eccentricities of the resonant pair is free to grow (and experience secular cycles) due to interactions with other extra planets in the system \citep[in the \thisstar system, there appear to be several additional planets; additionally, an increased eccentricity for either planet involved in the resonance may change the resonance width and lead eventually to the disruption of the resonance;][]{Wittenmyer:2012,Malhotra:2018}.

\citet{Beauge2003} presents solutions, inspired by Gliese-876, for stable aligned pericenters in the 2:1 resonance. Notably, stable solutions must have non-null eccentricities for the inner planet, and for an inner eccentricity of 0.3 or so, the outer planet must have a non-null eccentricity as well (for equal mass planets). The EXOFASTv2 posteriors indicate that \thisstarc could have a non-zero eccentricity; however, the pericenters are not well enough constrained to determine whether this aligned scenario occurs. 

Using our suite of N-body simulations, we can evaluate the fraction of fitted posteriors which are consistent with a resonant configuration for planets b and c. 
Of the 3000 trials considered, 32\% of them are dynamically stable. Of this stable subset, only one began in a true resonance (as defined by a librating resonant angle) and remained so for the entire integration. 
However, some of the integrations show that the planets can attain and lose a resonant configuration through their natural orbital evolution: in 7\% of the dynamically  stable integrations, \thisstar b and c attain a true 2:1 mean motion resonance for at least some of the integration (generally for periods around $10^{5}$ years at a time) but are subsequently disrupted from that resonance. A further 1\% attain resonance during the integration and remain stable in that resonance for the entire rest of the 1 Myr integration. However, 86\% of the dynamically  stable integrations never attain a resonant configuration. Barring one single integration that, after a scattering interaction, attained the 5:3 true resonance, the remaining $\sim$5\% of the stable simulations exhibit (for at least some of the integration, but not a majority) a `nodding' behavior \citep[i.e.,][]{Ketchum:2013} in and out of the 2:1 resonance. 

From this suite of simulations, it appears that the vast majority of the dynamically stable posteriors are fully non-resonant. 
However, a true orbital resonance could explain both the eccentricity of the inner planet and the continued stability of the system. 
The simulations show that this system, if in resonance, is likely characterized by non-consistent attainment of true resonance.

{\it 2) These planets formed in-situ or via scattering inwards, and do not have resonance in their history}. 
\citet{Terquem2018} note that only roughly 15\% of systems are consistent with smooth, disk-driven migration, which results in systems with $0 < \delta <$  0.04 (the 'outside of resonance' population common in the observational sample). If \thisstar was in resonance while the disk was still present (required if it assembled via disk-driven migration), it should have moved toward positive $\delta$ while in resonance and ended with orbits consistent with this population. Its small, negative value of $\delta$ can be explained if the system did not assemble via smooth migration, is not in resonance, and reached its current proximity to resonance by chance. 

{\it 3) These planets formed via disk-driven migration and were in resonance, but are no longer in resonance}.
As discussed in \citet{Adams:2008} and \citet{Batygin-Adams:2017}, turbulent fluctuations in the disk can destabilize resonances for small planets. These planets are both 2.7 Earth radii, slightly larger than should be possible to turbulently force out of resonance according to \citet{Batygin-Adams:2017}.
However, the 2:1 resonance is rather weak. As such, the resonant angles for this resonance generally have a large libration amplitude, potentially permitting either liberation from true resonance with minor perturbations, or large excursions in orbital element libration. 
As demonstrated by the numerical simulation, a sizable fraction of the posteriors attain and subsequently lose the 2:1 resonance during the integrations (sometimes multiple times).

\section{Discussion}
\label{sec:discussion}

Of the $4723$\ \Kepler\ planets and planet candidates discovered to date\footnote{Kepler Objects of Interest reported as Confirmed or Candidate planets on the NASA Exoplanet Archive on 2018 Oct 22.}, the majority are smaller than Neptune and larger than the Earth, and orbit within a few tenths of an AU \citep[e.g.,][]{thompson:2018}, a class of planet not seen in the Solar System. The occurrence rates of these short-period super-Earths and mini-Neptunes indicate that they are common byproducts of star formation \citep[e.g.,][]{Fressin:2013,Petigura:2013,Dressing:2015}. As such, their physical and orbital properties hold a wealth of information about the processes governing planet formation and evolution that were previously unconstrained by observation. The ensemble properties of these planets have revealed some of their fundamental characteristics \citep[e.g.,][]{Fulton:2017, Fulton:2018,berger:2018}, and detailed investigations of individual systems can complement the information gained from population studies.

\subsection{Radial Velocity Characterization}
\label{sec:rv}

The observed bimodality in the radius distribution of \kep\ planets \citep[e.g.,][]{Owen:2013,zeng:2017,Fulton:2017}, with peaks at $\mysim 1.3$\ and $2.4$\ \rearth, can be reproduced theoretically from the photoevaporation of close-in, low-mass planets, which are stripped to their bare ($\mysim 1.3$\ \rearth) cores, while more massive planets hold onto their H/He envelopes \citep[e.g.,][]{Owen:2017,Jin:2017}. If this were universally true, then the larger of these planets should be more or less the same mass as their smaller counterparts, as a H/He envelope will contribute a significant fraction of a planet's radius but very little mass. However, some planets with radii between $2$\ and $3$\ \rearth\ appear too dense for this scenario \citep[see, e.g., the recent \tess\ discovery of HD 21749 b;][]{dragomir:2019}. One explanation is that these denser sub-Neptune planets correspond to those in the large-core tail of the distribution. Another is that planet formation proceeds hierarchically, first accreting a rocky core, followed by CNO (e.g., water), and finally H/He, suggesting that planets of $\mysim2.4$\ \rearth\ correspond to ``water worlds''---planets with a high mean-molecular-weight envelope \citep{zeng:2017,zeng:2018}. However, this alternative would not explain the low-mass, large planets more consistent with an envelope of H/He. If both modes of planet formation operate, observation can constrain their relative occurrence. Systems like \thisstar, which contains three sub-Neptune planets of similar size and with a range of orbital periods, provide a good opportunity to measure the densities of these planets under controlled conditions. Having been subjected to the same stellar environment, conclusions based on the relative properties of the planets orbiting \thisstar\ are less affected by assumptions about stellar evolution and the history of stellar irradiation. As such, spectroscopic follow-up of \thisstar\ may provide insight into planet formation and evolution.

\thisstar\ is well suited to precise radial velocity measurements to determine the mass of its planets. The star is a bright ($V=11.0$; $T=10.1$), slowly rotating ($\vsini < 2$\,\kms), late-G dwarf with very little photometric variation ($\sigma_{\rm phot}$). The \citet{Chen:2017} planetary mass-radius relationship predicts masses of $8.5$, $8.6$, and $9.5\,\mearth$, corresponding to RV semi-amplitudes $3.7$, $2.9$, and $2.5$\,\ms. Given an instrumental precision of $\mysim 1$\,\ms\ for facilities such as HARPS and PFS, we expect that all three planets should have detectable RV signals.

\begin{figure*}[!ht]
\includegraphics[width=\linewidth]{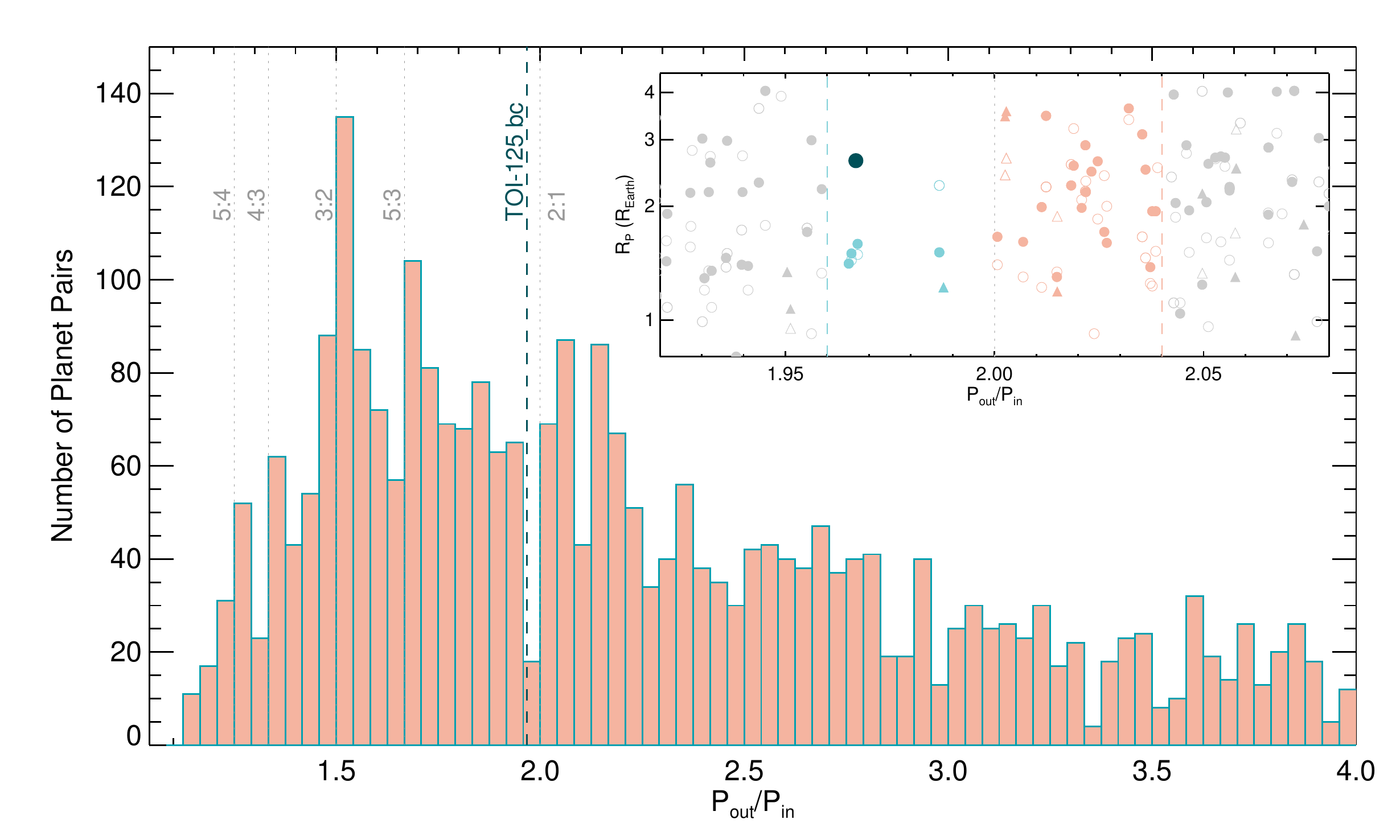}
\caption{Distribution of period ratios of all pairs of \kep\ candidates in multi-planet systems, excluding known false positives (orange histogram). Low-order resonances are shown in gray with dotted lines. The period ratio between \thisstarc and b is indicated by the vertical dashed blue line, and lies in the underpopulated region just short of 2:1. \textit{Inset}: Pairs of radii plotted against their period ratios near the 2:1 MMR. The excess of systems just wide of resonance (orange) and the dearth of systems just shy of resonance (blue) are apparent. Inner planets are shown with open symbols and outer planets with filled symbols. Circles and triangles represent adjacent and non-adjacent planet pairs, respectively. \thisstar is the larger dark blue circle; b and c lie on top of one another, and are the largest planets in the period ratio gap.}
\label{fig:prat}
\end{figure*}

We can estimate the time requirements to characterise the \thisstar\ system with the HARPS spectrograph using the RVFC tool developed by \citet{Cloutier:2018}. RV noise sources are estimated as a combination of the instrument noise floor ($0.$5\,\ms), the photon noise ($2.51$\,\ms\ for $30$-minute exposures), stellar activity ($0.5$\,\ms\ for a worst case \vsini\ of $2$\,\kms) and the RV rms caused by additional unseen planets (typically $0.4$\,\ms\ in this case). Details of how these noise sources are generated from the known stellar parameters, including Gaussian Process (GP) trials to simulate the stellar activity, are given in \citet{Cloutier:2018}. A complication for this system is its known multiplicity. Additional planets which are perfectly modelled do not impact the characterization of a planet, but given no model is perfect some additional rms will be present. We use the unseen planet RV rms estimate from RVFC as a zero-order guess at this contribution to the noise budget. We take the longest period planet as the driver of the necessary observations, implicitly assuming that observations sample well the  orbital phase curve of that and each interior planet, and that all planets are on circular orbits. Although photon noise dominates for this apparently low activity star, the effect of stellar activity can be large depending on the rotation period of the star, and in particular whether it is near a harmonic of the planet orbital periods. We present several representative cases, each calculated using the given stellar parameters, estimated planetary masses and $10$\ GP trials to estimate stellar activity. To characterize the system with a $5$-$\sigma$\ detection of the semi-amplitude for the outer planet (\thisstar.03) RVFC predicts $68\pm9$\ RV observations for a stellar rotation period of $25$\,days, rising to $141\pm37$\ observations for a difficult case rotation period of $40$\,days, double the orbital period of candidate .03. For the case of a $20$-day stellar rotation period characterizing candidate .03 becomes untenable, with RVFC predicting $45\pm5$\ observations to characterize only planets b and c. 

If real, the low SNR events \thisstarfour\ and \thisstarfive\ may complicate the RV detection of the other three planets, but they are also predicted to have detectable RV signals. With a timescale very different from the other planets and a predicted semi-amplitude of $2.2$\ \ms, \thisstarfour\ could be detected with a dedicated high-cadence RV campaign, which would ultimately benefit the detection of the outer signals, as it would otherwise enter as an additional source of noise. \thisstarfive\ is the lowest SNR candidate and least likely to prove real, but may also produce a detectable RV signal. Its predicted semi-amplitude given its derived size ($4.2$\ \rearth) is $\mysim 4.8$\ \ms. On the other hand, such a large planet is not typically seen in tightly packed systems. Therefore, we expect that if it is real, the planet is more likely to reside at the smaller end of the fit posteriors (see \rffigl{cand}), with an RV signal on the order of $1$--$2$\ \ms.

\subsection{\thisstar among the Kepler multis} 

Nearly $2000$\ of the $4723$\ \Kepler\ Objects of Interest (KOIs) reside in multi-planet systems, and their orbital architectures provide clues to their formation and evolution: the typical mutual inclination of short-period systems can be derived from the number of planets per system \citep[e.g.,][]{Lissauer:2011b,fang:2012,Ballard:2016}, and from the ratio of transit durations within each system \citep[e.g.,][]{fang:2012,Fabrycky:2014}, and inform the dynamical histories of planetary systems; the sizes and orbital spacing of neighboring planets relay information about formation and physical evolution \citep[e.g.,][]{weiss:2014}; the assembly of planets from planetesimals in the inner region of the protoplanetary disk can be examined through the lens of the present-day properties of short-period planets \citep[e.g.,][]{lee:2017}. 

One striking feature of the population of \kep\ multi-planet systems is the distribution of period ratios near first-order mean motion resonances. As discussed in \rfsecl{dynamics}, there is an underdensity of planet pairs just interior to first-order resonances---particularly the 2:1 resonance---and an excess of systems just exterior to resonance. We present an updated histogram of period ratios for \kep\ systems in \rffigl{prat} \citep[see][ for a broader discussion of these data]{Steffen:2013,Fabrycky:2014}, and we note that \thisstarb\ and c have a period ratio that falls right in the gap interior to the 2:1 resonance. We have explored possible causes for this in \rfsecl{dynamics}, and here we compare \thisstar\ to the small handful of other systems in or near this gap, seen in the inset of \rffigl{prat}. 

None of the five other systems interior to, but within $2\%$ of 2:1, is quite like \thisstar, which is larger and/or shorter period than the others.  Kepler-176\,d (KOI-520.03) is the most similar in size to \thisstarb\ and c, but it, along with Kepler-334 (KOI-1909), is longer period \citep[weeks, rather than days;][]{rowe:2014}. Kepler-271 (KOI-1151) and KOI-4504 have similar periods to \thisstar, but the planets are much smaller. One other system (KOI-1681) has an interesting architecture, with one small planet and one hot Jupiter \citep[similar to WASP-47;][]{Becker:2015}, but it also has a nearby stellar companion, a third signal that is likely a false positive, and a fourth signal corresponding to another giant planet candidate. If real, it would be unlike any other system that we know. \thisstar\ is thus a unique system even within its sparsely populated region of parameter space, and worthy of additional study. 

\subsection{The Candidate USP Planet \thisstarfour}

While \thisstarfour\ was only detected with ${\rm SNR}\mysim 5.2$, the presence of three high-SNR transit signals in the system makes it more likely that the signal is real compared to an isolated signal of similar strength. Moreover, the architecture of the \thisstar\ planets would match that of other known USP systems, both in semi-major axis and in mutual inclination. \citet{dai:2018} find that \kep\ and \Ktwo\ planets in multi-planet systems tend to exhibit large mutual inclinations when the innermost planet has a small semi-major axis ($a/R_* < 5)$, and that among these USP planets, the systems with the largest mutual inclinations also have large period ratios \citep[see, e.g., K2-266, with an extreme mutual inclination of $\mysim12$\ degrees][]{rodriguez:2018b}. \thisstarfour\ orbits at ($a/R_*\mysim3.1$) and with a projected mutual inclination of $\mysim16$\ degrees. The period ratio between \thisstarb\ and \thisstarfour\ is $8.8$, similar to the other such misaligned USP planets. \thisstarfour\ would be the USP planet with the largest known mutual inclination, but as discussed above, \thisstarb\ is larger than most inner planets in packed systems. We speculate that in the framework suggested by \citet{dai:2018}, one would expect the interaction between the two planets that leads to the inclined USP planet to produce a more extreme outcome when the adjacent planet is more massive than usual.

\subsection{Dynamical Results}
The EXOFASTv2 transit fit allowed a relatively large range in eccentricities for \thisstar b, c, and d, but the largest of the allowed values can be excluded due to dynamical stability constraints. 
Of the entire EXOFASTv2 posteriors, approximately 32\% of draws result in integrations that remain dynamically stable for 1 Myr. 
Preferentially, the stable subset are those with lower eccentricities: eccentricities above 0.25--0.3 are disallowed for each planet (see Figure \ref{sec:stability}). 

Using the results of the numerical simulations to inform the resonant behavior paints a largely non-resonant picture of the posteriors. 
Of the stable subset of integrations, roughly 86\% exhibited nearly exclusively non-resonant behavior for planets b and c. In these cases, the median period ratio libration has a (min-to-max) amplitude of $\Delta\delta \approx 0.01$, surrounding a median $\delta$ value of $\delta \approx -0.032 $ (compared to the measured current-day value of $\delta \approx -0.033 $).
From these simulations, it is reasonable that the observed three planet system can remain dynamically stable in the observed orbits for a relatively large fraction of the EXOFASTv2 posteriors. 
In comparison, among the $\sim12\%$ of integrations that reside in or nod in/out of the 2:1 resonance for some or all of of the 1 Myr, while in resonance the size of $\Delta\delta$ depends on the libration width of the resonant angle. For nodding, this value is as large as $\approx 0.08$, around a median $\delta$ value of $\delta \approx 0.001$. In true resonance, the typical $\Delta\delta$ values range between 0.02 and 0.04, as the median $\delta$ value becomes closer to $\delta \approx 0$.
It is possible that \thisstar b and c previously resided in the 2:1 resonance and naturally lost the resonance at some point, becoming trapped in a dynamically stable but non-resonant orbit
Although the errors on the currently measured orbit are too large to conclusively determine the current-day resonance behavior of \thisstar b and c, the simulations suggest they are most likely not currently in resonance, but have eccentricities near the lower end of the measured posteriors. 

\section{Summary}
\label{sec:summary}

In this paper, we have presented the \tess\ discovery of the \thisstar\ multi-planet system, and we fit a global model to the \tess\ data, spectroscopic stellar parameters, literature photometry, and the \textit{Gaia} parallax in conjunction with stellar models to characterize the planet candidates. We then statistically validated the planetary nature of \thisstarb and c using \vespa\ with the aid of  archival imaging and our photometric, spectroscopic, and high resolution imaging observations. We demonstrated that the system is likely amenable to mass determination via both TTV and precise RV follow-up, and that the planets are worthy of such additional study. The three strongest transit signals are caused by planets with radii $2.8$--$2.9$\,\rearth, a class of planet not seen in the Solar System but abundant in the galaxy. These planets have been proposed as the progenitors (via photoevaporation) of the terrestrial planets commonly found in short periods around nearby stars, and studying three of them in the controlled environment of the same host star can help illuminate the formation and evolution processes at play. The candidate terrestrial USP planet, with an orbital period less than $13$\ hours and a mutual inclination of $16$\ degrees with the other planets, is an extreme example of the trend toward such architectures among other known USP planets in multiple systems, and may be the end result of dynamical interaction with its much larger sub-Neptune neighbors. Finally, the period ratio between planets b and c is very near, but just interior to, a 2:1 commensurability, which is quite unusual compared to known \kep\ systems. While one possible explanation is that the system is in---and librating about---the 2:1 resonance, our dynamical analysis suggests that it is unlikely that the system is currently in true resonance. The discovery of the \thisstar\ system demonstrates that \tess\ continues in its early days to deliver on its promise to identify rare systems of small planets amenable to follow-up observations and detailed characterization.

\acknowledgments

JCB is supported by the NSF Graduate Research Fellowship Grant No. DGE 1256260 and by the Leinweber Center for Theoretical Physics. Work performed by JER was supported by the Harvard Future Faculty Leaders Postdoctoral fellowship. DJA gratefully acknowledges support from the STFC via an Ernest Rutherford Fellowship (ST/R00384X/1). AV's contribution to this work was performed under contract with the California Institute of Technology (Caltech)/Jet Propulsion Laboratory (JPL) funded by NASA through the Sagan Fellowship Program executed by the NASA Exoplanet Science Institute. Work by JNW was supported by the Heising-Simons Foundation. DD acknowledges support for this work provided by NASA through Hubble Fellowship grant HST-HF2-51372.001-A awarded by the Space Telescope Science Institute, which is operated by the Association of Universities for Research in Astronomy, Inc., for NASA, under contract NAS5-26555. AG is supported by the Ida M. Green Fellowship. M.T. is supported by MEXT/JSPS KAKENHI grant Nos. 18H05442, 15H02063, 
and 22000005. This work is partly supported by JSPS KAKENHI Grant Numbers JP15H02063, JP18H01265, JP18H05439, JP18H05442, and JST PRESTO Grant Number JPMJPR1775. We acknowledge the support provided by the Polish National Science 
Center through grants 2016/21/B/ST9/01613 and 2017/27/B/ST9/02727.

We thank Zach Hartman, Dan Nusdeo, and Jen Winters for help with the Gemini-South observations. We thank Akihiko Fukui, Nobuhiko Kusakabe, Kumiko Morihana, Tetsuya Nagata, Takahiro Nagayama, Taku Nishiumi, and the staff of SAAO for their kind support for IRSF SIRIUS observations and analyses.

Funding for the \tess\ mission is provided by NASA's Science Mission directorate. We acknowledge the use of public \tess\ Alert data from pipelines at the \tess\ Science Office and at the \tess\ Science Processing Operations Center. This paper includes data collected by the \tess\ mission, which are publicly available from the Mikulski Archive for Space Telescopes (MAST).

This work used the Extreme Science and Engineering Discovery Environment (XSEDE), which is supported by National Science Foundation grant number ACI-1053575. This research was done using resources provided by the Open Science Grid, which is supported by the National Science Foundation and the U.S. Department of Energy's Office of Science, through allocation  TG-AST150033.

The Swiss Euler telescope is supported by the Swiss National Science Foundation. This paper includes observations obtained under ESO/VLT program 0102.C-0503(A), and Gemini program GS-2018B-LP-101. Gemini Observatory is operated by the Association of Universities for Research in Astronomy, Inc., under a cooperative agreement with the NSF on behalf of the Gemini partnership: the National Science Foundation (United States), National Research Council (Canada), CONICYT (Chile), Ministerio de Ciencia, Tecnolog\'{i}a e Innovaci\'{o}n Productiva (Argentina), Minist\'{e}rio da Ci\^{e}ncia, Tecnologia e Inova\c{c}\~{a}o (Brazil), and Korea Astronomy and Space Science Institute (Republic of Korea). This work makes use of observations from the LCOGT network. The IRSF project is a collaboration between Nagoya University and the South African Astronomical Observatory (SAAO) supported by the Grants-in-Aid for Scientific Research on Priority Areas (A) (Nos. 10147207 and 10147214) and Optical \& Near-Infrared Astronomy Inter-University Cooperation Program, from the Ministry of Education, Culture, Sports, Science and Technology (MEXT) of Japan and the National Research Foundation (NRF) of South Africa. Based in part on data collected with {\it Solaris} network of telescopes of the Nicolaus Copernicus Astronomical Center of the Polish Academy of Sciences. The research leading to these results has received funding from the European Research Council under the European Union's Seventh Framework Programme (FP/2007-2013) ERC Grant Agreement 336480, from the ARC grant for Concerted Research Actions, financed by the Wallonia-Brussels Federation, and from a research grant from the Balzan Prize Foundation. TRAPPIST is funded by the Belgian Fund for Scientific Research (Fond National de la Recherche Scientifique, FNRS) under the grant FRFC 2.5.594.09.F, with the participation of the Swiss National Science Fundation (SNF). M.G. and E.J. are FNRS Senior Research Associates.

This work has made use of data from the European Space Agency (ESA) mission {\it Gaia} (\url{https://www.cosmos.esa.int/gaia}), processed by the {\it Gaia} Data Processing and Analysis Consortium (DPAC, \url{https://www.cosmos.esa.int/web/gaia/dpac/consortium}). Funding for the DPAC has been provided by national institutions, in particular the institutions participating in the {\it Gaia} Multilateral Agreement. This research has made use of NASA's Astrophysics Data System. This research has made use of the VizieR catalogue access tool, CDS, Strasbourg, France. The original description of the VizieR service was published in A\&AS 143, 23. This research has made use of the SIMBAD database, operated at CDS, Strasbourg, France. This research has made use of the NASA Exoplanet Archive and the Exoplanet Follow-up Observation Program website, which are operated by the California Institute of Technology, under contract with the National Aeronautics and Space Administration under the Exoplanet Exploration Program.

\facilities{TESS, Euler1.2m (CORALIE), Gemini:South (DSSI), SOAR (HRCam), VLT:Antu (NACO), IRSF, LCOGT, SSO:1m, TRAPPIST}

\software{\texttt{EXOFASTv2} \citep{Eastman:2013,Eastman:2017}, \vespa\ \citep{morton:2015}, \texttt{Mercury6} \citep{Chambers:1999}, \texttt{AstroImageJ} \citep{collins:2017}}

\bibliographystyle{apj}

\bibliography{toi125}

\allauthors
\end{document}